\documentstyle[prd,aps]{revtex}

\input epsf
\newcommand{\bea}{\begin{eqnarray}}
\newcommand{\eea}{\end{eqnarray}}

\begin{document}
\twocolumn[\hsize\textwidth\columnwidth\hsize\csname
@twocolumnfalse\endcsname

\setcounter{page}{1}
\title{Gauge-ready formulation of the cosmological kinetic theory \\
       in generalized gravity theories}
\author{Jai-chan Hwang${}^{(a,b)}$ and Hyerim Noh${}^{(c,b)}$ \\
        ${}^{(a)}$ Department of Astronomy and Atmospheric Sciences,
                   Kyungpook National University, Taegu, Korea \\
        ${}^{(b)}$ Institute of Astronomy, Madingley Road, Cambridge, UK \\
        ${}^{(c)}$ Korea Astronomy Observatory, Taejon, Korea}
\date{\today}
\maketitle

\begin{abstract}

We present cosmological perturbations of kinetic components based on 
relativistic Boltzmann equations in the context of generalized gravity 
theories.
Our general theory considers an arbitrary number of scalar fields generally 
coupled with the gravity, an arbitrary number of mutually interacting 
hydrodynamic fluids, and components described by the relativistic Boltzmann 
equations like massive/massless collisionless particles and the photon
with the accompanying polarizations.
We also include direct interactions among fluids and fields.
The background FLRW model includes the general spatial curvature and the 
cosmological constant.
We consider three different types of perturbations, and all the scalar-type 
perturbation equations are arranged in a gauge-ready form so that one can 
implement easily the convenient gauge conditions depending on the situation.
In the numerical calculation of the Boltzmann equations we have implemented 
four different gauge conditions in a gauge-ready manner where two of them 
are new.
By comparing solutions solved separately in different gauge conditions
we can naturally check the numerical accuracy.

\end{abstract}


\vskip2pc]

\section{Introduction}
                                          \label{sec:Introduction}

The relativistic cosmological perturbation plays a fundamental role
in the modern theory of large-scale cosmic structure formation based
on the gravitational instability.
Due to the extremely low level anisotropies of the cosmic microwave 
background (CMB) radiation, the cosmological dynamics of the structures 
in the large-scale and in the early universe are generally believed to be 
operated as small
deviations from the homogeneous and isotropic background world model.
The relativisitic cosmological perturbation analysis works as the basic
framework in handling such cosmological structure formation processes.
Recent observations of the CMB anisotropies in the small-angular scale
by Boomerang and Maxima-1 experiments 
\cite{deBernardis-etal-2000,Hanany-etal-2000}, for example, confirm 
dramatically the validity of the basic assumptions used in the
cosmological perturbation theory, i.e., the linearity of the relevant
cosmic structures.

Soon after the discovery of the CMB by Penzias and Wilson in 1965
\cite{Penzias-Wilson-1965}, Sachs and Wolfe in 1967 \cite{Sachs-Wolfe-1967} 
pointed out that the CMB should show the 
temperature anisotropy caused by photons traveling in the perturbed metric 
which is associated with the large-scale structure formation processes 
based on the gravitational instability.
The detailed dynamics at last scattering is not important in the large 
angular scale which can be handled using the null geodesic equations.
Whereas, the physical processes of last scattering including the
recombination process are important in the small angular scale where 
we need to solve the Boltzmann equations for the
photon distribution function \cite{Peebles-Yu-1970}. 
When we handle the evolutions of the collisionless particles,
like the massive/massless neutrinos or the collisionless dark matters,
we need the corresponding Boltzmann equations as well.

The relativistic gravity theory, including Einstein's general theory of
relativity as a case, is a non-Abelian gauge theory of a special type.
The original perturbation analysis was made by Lifshitz in 1946 based on 
Einstein gravity with the hydrodynamic fluid \cite{Lifshitz-1946}.
In handling the gauge degrees of freedom arising in the perturbation
analyses in the relativistic gravity, Lifshitz started by choosing the
synchronous gauge condition and properly sorted out the remaining 
gauge degrees of freedom incompletely fixed by his gauge condition.
Other approaches based on other (more suitable) gauge conditions
were taken by Harrison using the zero-shear gauge in 1967 \cite{Harrison-1967} 
and by Nariai using the comoving gauge in 1969 \cite{Nariai-1969}. 
Each of these two gauge conditions completely removes the gauge degrees 
of freedom.
Now, we know that the zero-shear gauge is suitable for handling the 
gravitational potential perturbation and the velocity perturbation, 
and the comoving gauge is suitable for handling the density perturbation.
Since each of these two gauge conditions completely fixes the gauge
transformation properties, all the variables in the gauge condition are
the same as the gauge-invariant ones: that is, each variable uniquely
corresponds to a gauge-invariant combination of the variable concerned
and the variable used in the gauge condition.

The gauge-invariant combinations were explicitly introduced by Bardeen 
in 1980 \cite{Bardeen-1980}; see also Lukash 1980 \cite{Lukash-1980} 
for a parallelly  important contribution.
This became a seminal work due to a timely introduction of the early 
inflation scenario \cite{Inflation}
which provides a casual mechanism for explaining the generation and 
evolution of the observed large-scale cosmic structures.
We believe, however, a practically more important suggestion concerning 
the gauge issue was made by Bardeen in 1988 \cite{Bardeen-1988},  
and it was elaborated in \cite{Hwang-1991-PRW}. 
In the gauge theory it is well known that proper choice of the gauge
condition is often necessary for proper handling of the problem.
Either by fixing certain gauge conditions or by choosing certain
gauge-invariant combinations in the early calculation stage we are likely to 
lose possible advantages available in the other gauge conditions.
According to Bardeen ``the moral is that one should work in the gauge
that is mathematically most convenient for the problem at hand''.
In order to use the various gauge conditions as advantages in handling 
cosmological perturbations we have proposed a gauge-ready method
which allows the flexible use of the various fundamental gauge conditions.
In this paper we will elaborate further the gauge-ready approach for more 
general situations of the generalized gravity theories including components 
described by the relativistic Boltzmann equations.

Our formulation is made based on the gauge-ready approach; using this 
approach our new formulation of the cosmological perturbation is more 
flexible and adaptable in practical applications compared with previous works.
Also, the formulation is made for a Lagrangian which is very general,
thus includes most of the practically interesting generalized versions
of gravity theories considered in the literature.
We pay particular attention to make the contribution of
the kinetic components in the context of the generalized gravity theories.
As an application of our gauge-ready approach made in this paper,
we implemented the numerical integration of Boltzmann equations for CMB 
anisotropies in four different gauge conditions.
In addition to the previously used synchronous gauge (without the gauge mode)
and the zero-shear gauge, we also implemented
the uniform-expansion gauge and the uniform-curvature gauge
in a gauge-ready manner.
These two gauge conditions have not been employed in the study of the
CMB power spectra previously. 
We will show that by comparing solutions solved separately in different
gauge conditions we can naturally check the numerical accuracy.

In \S \ref{sec:Equations} we present the classical formulation of
the cosmological perturbations of fields and fluids in the context of
generalized gravity in a unified manner;
i.e., diverse gravity theories are handled in a unified form.
The formulation is based on the gauge-ready strategy which is explained
thoroughly in \S \ref{sec:gauge}.
In \S \ref{sec:Kinetic} we present the gauge-ready formulation of the kinetic
components based on the relativistic Boltzmann equations in the context
of generalized gravity again in unified manner;
i.e., we handle the massive/massless collisionless particles and the
photon with Thomson scattering simultaneously, and all three-types of
perturbations are handled in a single set of equations.
In \S \ref{sec:CMB} we extend the formulation to include the photon with 
polarizations, and implement the numerical calculation of the CMB
temperature and polarization anisotropy power spectra. 
Our present code is based on Einstein gravity
including the baryon, cold dark matter (CDM), photon (including polarizations), 
massless/massive neutrinos, the cosmological constant, and the
background curvature, both for the scalar- and tensor-type perturbations.
The scalar-type perturbation is implemened using several gauge conditions 
where some of them are new. 
We explain how to generalize easily the Boltzmann code in the context of
the generalized gravity theories including recently popular time varying 
cosmological constant.
\S \ref{sec:Discussion} is a discussion.
In the Appendices \ref{sec:CT} and \ref{sec:Fluid-quantities}
we present the conformal transformation properties
of our generalized gravity theories and the effective fluid quantities.
In the Appendix \ref{sec:Kinematic} we present useful kinematic quantities 
appearing in the $3+1$ ADM (Arnowitt-Deser-Misner) formulation and
the $1+3$ covariant formulation of the cosmological perturbation theory.

We set $c \equiv 1$.

\section{Classical Formulation}
                                          \label{sec:Equations}

\subsection{Generalized gravity theories}
                                          \label{sec:GGT}

We consider a gravity with an arbitrary number of scalar fields 
generally coupled with the gravity, and with an arbitrary number 
of mutually interacting imperfect fluids as well as the kinetic components.
As the Lagrangian we consider
\bea
   {\cal L} 
   &=& \sqrt{-g}
       \Big[ {1 \over 2} f (\phi^K, R)
       - {1\over 2} g_{IJ} (\phi^K) \phi^{I;c} \phi^J_{\;\;,c}
       - V (\phi^K) 
   \nonumber \\ 
   & & + L_m \Big].
   \label{Lagrangian}
\eea
$R$ is the scalar curvature. 
$\phi^I$ is the $I$-th component of $N$ scalar fields.
The capital indices $I, J, K, \dots = 1,2,3,$ $\dots, N$ indicate the 
scalar fields, and the summation convention is used for repeated indices.
$g \equiv {\rm det}(g_{ab})$ where $a,b, \dots $ are spacetime indices.
$f(\phi^K,R)$ is a general algebraic function of $R$ and the scalar
fields $\phi^I$, and $g_{IJ} (\phi^K)$ and $V(\phi^K)$ are general
algebraic functions of the scalar fields;
$f(\phi^K,R)$ and $V(\phi^K)$ indicate
$f(\phi^1, \dots, \phi^N, R)$ and $V (\phi^1, \dots , \phi^N)$.
We include a nonlinear sigma type kinetic term where
the kinetic matrix $g_{IJ}$ is considered as a Riemannian metric on the
manifold with the coordinates $\phi^I$.
The matter part Lagrangian $L_m$ includes the fluids,
the kinetic components, and the interaction with the fields, as well.

Equation (\ref{Lagrangian}) contains many interesting gravity theories
with scalar fields as subsets.
Einstein gravity is a case of the minimal coupling with the gravity, 
thus $f = R/(8 \pi G)$;
this case still includes the nonlinear sigma type couplings among fields, 
and for the minimally coupled scalar fields we have $g_{IJ} = \delta_{IJ}$.
The general couplings of the scalar fields with gravity and the
nonlinear sigma type kinetic term generically appear in various
attempts to unify the gravity with other fundamental forces, like
the Kaluza-Klein, the supergravity, the superstring, and the M-theory programs; 
these terms also appear naturally in the quantization processes of the gravity
theory in a way toward the quantum gravity.
The Lagrangian in eq. (\ref{Lagrangian}) includes the following generalized
gravity theories as subsets [for simplicity, we consider one scalar field 
with $\phi \equiv \phi^1$ and $g_{IJ} = g_{11} (\phi)$]:

   (a) Einstein theory:
       $f = {1 \over 8 \pi G} R, \; g_{11} = 1$,

   (b) Brans-Dicke theory:
       $f = {1 \over 8 \pi} \phi R, \;
       g_{11} = {\omega \over 8 \pi \phi}, \; V = 0$,

   (c) Low-energy string theory:

       $\qquad f = e^{-\phi} R, \; g_{11} = - e^{-\phi}, \; V = 0$,

   (d) Nonminimally coupled scalar field:

       $\qquad f = ({1 \over 8 \pi G} -\xi \phi^2) R, \; g_{11} = 1$,

   (e) Induced gravity:

       $\qquad f = \epsilon \phi^2 R, \; g_{11} = 1, \;
       V = {1 \over 4} \lambda (\phi^2 - v^2 )^2$,

   (f) $R^2$ gravity:
       $f = {1 \over 8 \pi G} ( R + R^2/6M^2), \; \phi = 0$,
\bea
   \label{gravities}
\eea
etc.
These gravity theories without additional fields and matters can be considered
as the second-order theories.
However, even with a single scalar field, the $f(\phi,R)$ gravity 
is generally a fourth-order theory.
Although such gravity theories do not have an immediate interest in 
the context of currently considered generalized gravity theories, 
one simple example is a case with $f = f_1(\phi) f_2(R)$ where $f_2 (R)$ is a
nonlinear function of $R$.

By the conformal transformation eq. (\ref{Lagrangian}) can be transformed
to Einstein gravity with nonlinear sigma model type scalar fields,
and the transformed theory also belongs to the type in eq. (\ref{Lagrangian});
see the Appendix \ref{sec:CT}.
Authors of \cite{Salopek-etal-1989} considered a less general 
form of Lagrangian than in eq. (\ref{Lagrangian}) in perturbation analyses;
however, since they used the conformal transformation, 
they actually considered Einstein gravity with nonlinear sigma type couplings. 

Variations with respect to $g_{ab}$ and $\phi^I$ lead to 
the gravitational field equation and the equations of motion:
\bea
   & & G_{ab} = {1 \over F} \Bigg[ T_{ab}
       + g_{IJ} \left( \phi^I_{\;\;,a} \phi^J_{\;\;,b}
       - {1 \over 2} g_{ab} \phi^{I;c} \phi^J_{\;\;,c} \right)
   \nonumber \\
   & & \quad
       + {1 \over 2} \left( f - RF - 2 V \right) g_{ab}
       + F_{,a;b} - g_{ab} F^{;c}_{\;\;\; c} \Bigg] 
   \nonumber \\
   & & \quad
       \equiv 8 \pi G T_{ab}^{({\rm eff})},
   \label{GFE} \\
   & & \phi^{I;c}_{\;\;\;\; c} + {1 \over 2} \left( f - 2 V \right)^{;I}
       + \Gamma^I_{JK} \phi^{J;c} \phi^K_{\;\;,c} 
       = - L_m^{\;\;\; ;I} \equiv \Gamma^I,
   \nonumber \\
   \label{EOM} \\
   & & T_{a;b}^{b} = L_{m,J} \phi^J_{\;\; ,a},
   \label{fluid-conservation}
\eea
where $F \equiv \partial f / \partial R$;
$g^{IJ}$ is the inverse metric of $g_{IJ}$, 
$\Gamma^I_{JK} \equiv {1 \over 2} g^{IL} 
\left( g_{LJ,K} + g_{LK,J} - g_{JK,L} \right)$,
and $V_{,I} \equiv \partial V/(\partial \phi^I)$.
Equation (\ref{fluid-conservation}) follows from 
eqs. (\ref{GFE},\ref{EOM}) and the Bianchi identity.
$T_{ab}$ is the energy-momentum tensor of the matter part defined as
$\delta ( \sqrt{-g} L_m )$ 
$\equiv$ ${1 \over 2} \sqrt{-g} T^{ab} \delta g_{ab}$.
We have assumed the matter part Lagrangian $L_m$ also depends on
the scalar fields as $L_m = L_m ({\rm matter}, g_{ab}, \phi^K)$.
In eq. (\ref{EOM}) $\Gamma^I$ term considers the phenomenological couplings
among the scalar fields and the matter.
In eq. (\ref{GFE}) we introduced an effective energy-momentum tensor
$T_{ab}^{({\rm eff})}$ where the matter $T_{ab}$
includes the fluids and the kinetic components.
The effective fluid quantities to the perturbed order are
presented in the Appendix \ref{sec:Fluid-quantities}.
Using $T_{ab}^{({\rm eff})}$ we can derive the fundamental
cosmological equations in the generalized gravity without much algebra: 
we use the same equations derived in Einstein gravity with the fluid 
energy-momentum tensor and reinterprete the fluid quantities 
as the effective ones \cite{Hwang-GGT-1990}. 
Direct derivation is also straightforward.

The matter energy-momentum tensor can be decomposed covariantly into
the fluid quantities using a normalized ($u^a u_a \equiv -1$) four-vector $u_a$ 
which is not necessarily the flow four-vector \cite{covariant}: 
\bea
   & & T_{ab} = \mu u_a u_b + p h_{ab} + q_a u_b + q_b u_a + \pi_{ab},
   \nonumber \\
   & & \mu \equiv T_{ab} u^a u^b, \quad
       p \equiv {1 \over 3} T_{ab} h^{ab}, \quad
       q_a \equiv - T_{cd} u^c h_a^d, \quad
   \nonumber \\
   & & \pi_{ab} \equiv T_{cd} h_a^c h_b^d - p h_{ab}, 
   \label{Tab-f}
\eea
where $h_{ab} \equiv g_{ab} + u_a u_b$ is a projection tensor of the $u_a$ 
vector, and $q_a u^a = 0 = \pi_{ab} u^b$,
$\pi_{ab} = \pi_{ba}$, and $\pi^a_a = 0$.
The matter energy-momentum tensor can be decomposed into the sum of the 
individual one as
\bea
   & & T_{ab} = \sum_{l} T_{(l)ab}, 
   \label{Tab-sum} 
\eea
and the energy-momentum conservation gives
\bea
   & & T^{\;\;\;\; b}_{(i)a;b} \equiv Q_{(i)a}, \quad
       \sum_{l} Q_{(l)a} = - \Gamma_I \phi^I_{\;,a}, 
   \label{Tab-i-conservation}
\eea
where $(i)$ indicates the $i$-th component of $n$ matters with
$i,j,k, \dots = 1,2,3,\dots n$.
The matters include not only the general imperfect fluids, but also
the contributions from multiple components of
the collisionless particles and the photon described by the 
corresponding distribution functions and the Boltzmann equations.
These kinetic components will be considered in 
\S \ref{sec:Kinetic} and \ref{sec:CMB}.
$Q_{(i)a}$ takes into account of possible interactions among matters and fields.

\subsection{Perturbed world model}
                                          \label{sec:pert-model}

We consider the most general perturbations in the 
FLRW (Friedmann-Lema\^itre-Robertson-Walker) world model.
As the metric we take
\bea
   d s^2 
   &=& - a^2 \left( 1 + 2 A \right) d \eta^2 
       - 2 a^2 B_\alpha d \eta d x^\alpha
   \nonumber \\
   & & + a^2 ( g^{(3)}_{\alpha\beta} 
       + 2 C_{\alpha\beta} ) d x^\alpha d x^\beta,
   \label{metric-general}
\eea
where $a(t)$ is the cosmic scale factor and $dt \equiv a d \eta$.
$A({\bf x}, t)$, $B_\alpha ({\bf x},t)$, and $C_{\alpha\beta} ({\bf x}, t)$
are generally spacetime dependent perturbed  order variables.
$B_\alpha$, $C_\alpha$, and $C_{\alpha\beta}$ are based on 
$g^{(3)}_{\alpha\beta}$, i.e., indices are raised and lowered
with $g^{(3)}_{\alpha\beta}$.

The scalar fields are decomposed into the background and perturbed parts as
\bea
   & & \phi^I ({\bf x}, t) = \bar \phi^I (t) + \delta \phi^I ({\bf x}, t),
   \label{phi-pert}
\eea
and similarly for $R$ and $F$.
In the following, unless necessary, we neglect the overbars which indicate
the background order quantities.

The energy-momentum tensor is decomposed as
\bea
   & & T^0_0 = - \mu \equiv - ( \bar \mu + \delta \mu ),
   \nonumber \\
   & & T^0_\alpha = {1 \over a} [ q_\alpha + ( \mu + p ) u_\alpha ]
       \equiv ( \mu + p ) v_\alpha,
   \nonumber \\
   & & T^\alpha_\beta = p \delta^\alpha_\beta + \pi^\alpha_\beta
       \equiv ( \bar p + \delta p ) \delta^\alpha_\beta 
       + \pi^{(3)\alpha}_{\;\;\;\;\;\beta},
   \label{Tab-decomp}
\eea
where $v_\alpha$ and $\pi^{(3)\alpha}_{\;\;\;\;\;\beta}$ 
are based on $g_{\alpha\beta}^{(3)}$.
$v_\alpha$ is a frame-independent definition of the velocity
(or the flux related) variable \cite{Hwang-1991-PRW}.
In the multi-component fluid situation from eq. (\ref{Tab-sum}) we have
\bea
   & & \bar \mu = \sum_{l} \bar \mu_{(l)}, \quad
       \delta \mu = \sum_{l} \delta \mu_{(l)},
\eea
and similarly for $\bar p$, $\delta p$, $(\mu + p) v_\alpha$, and
$\pi^{(3)\alpha}_{\;\;\;\;\;\beta}$.

\subsection{Decompositions}
                                          \label{sec:decomposition}

In the spatially homogeneous and isotropic background we can decompose
the perturbed variables into three different types, and to the linear order 
different perturbation types decouple from each other and evolve independently.
We decompose the metric perturbation variables $A$, $B_\alpha$, 
and $C_{\alpha\beta}$ as
\bea
   & & A \equiv \alpha,
   \nonumber \\
   & & B_\alpha \equiv \beta_{,\alpha} + B^{(v)}_\alpha,
   \nonumber \\
   & & C_{\alpha\beta} \equiv g^{(3)}_{\alpha\beta} \varphi
       + \gamma_{,\alpha|\beta}
       + C^{(v)}_{(\alpha|\beta)} + C^{(t)}_{\alpha\beta},
   \label{metric-decomp}
\eea
where $(s)$, $(v)$ and $(t)$ indicate the scalar-, vector- and tensor-type
perturbations, respectively.
$B^{(v)}_\alpha$, $C^{(v)}_\alpha$ and $C^{(t)}_{\alpha\beta}$
are based on $g^{(3)}_{\alpha\beta}$, and a vertical bar indicates 
a covariant derivative based on $g^{(3)}_{\alpha\beta}$;
for $t_{(\alpha\beta)}$ symbol see below eq. (\ref{kinematic-n-def}).
The perturbed order variables $\alpha ({\bf x}, t)$, $\beta ({\bf x}, t)$, 
$\varphi ({\bf x}, t)$, and $\gamma ({\bf x}, t)$ 
are the scalar-type metric perturbations.
$B^{(v)}_{\alpha} ({\bf x}, t)$ and $C^{(v)}_{\alpha} ({\bf x}, t)$ are 
transverse ($B^{(v)\alpha}_{\;\;\;\;\;\;\;|\alpha} = 0 
= C^{(v)\alpha}_{\;\;\;\;\;\;\;|\alpha}$)
vector-type perturbations corresponding to the rotational perturbation.
$C^{(t)}_{\alpha\beta} ({\bf x}, t)$ is a transverse-tracefree
($C^{(t)\alpha}_{\;\;\;\;\alpha} = 0 = C^{(t)\beta}_{\;\;\;\;\;\alpha|\beta}$)
tensor-type perturbation corresponding to the gravitational wave.
Thus, we have four degrees of freedom for the scalar-type, 
four degrees of freedom for the vector-type, 
and two degrees of freedom for the tensor-type perturbations.
Two degrees of freedom for the tensor-type perturbation 
indicate the graviational wave.
Whereas, two out of four degrees of freedoms, each for the scalar-type
and vector-type perturbations, are affected by coordinate
transformations which connect the physical perturbed spacetime
with the fictitious background spacetime.
This is often called the gauge effect and a way of using it as
{\it advantages} in handling problems will be described in 
\S \ref{sec:gauge}.
It is convenient to introduce the following combinations of the
metric variables:
\bea
   & & \chi \equiv a \left( \beta + a \dot \gamma \right), \quad
       \kappa \equiv 3 \left( H \alpha - \dot \varphi \right)
       - {\Delta \over a^2} \chi,
   \nonumber \\
   & & \Psi^{(v)} \equiv b^{(v)} + a \dot c^{(v)},
   \label{chi-kappa}
\eea
where an overdot indicates a time derivative based on $t$, and
$H \equiv \dot a/a$; $\Delta$ is a comoving three-space Laplacian, i.e., 
$\Delta \chi \equiv \chi^{|\alpha}_{\;\;\;\alpha}$.
Later we will see that these combinations are spatially gauge-invariant.
The perturbed metric variables have clear meaning based on the
kinematic quantities of the normal-frame four-vector,
see eqs. (\ref{kinematic-n},\ref{ADM-variables}).

We introduce three-space harmonic functions depending on the perturbation types.
The harmonic functions based on $g^{(3)}_{\alpha\beta}$ are introduced 
in \cite{Bardeen-1980,Kodama-Sasaki-1984}:
\bea
   & & Y^{(s)|\gamma}_{\;\;\;\;\;\;\;\;\;\gamma} \equiv - k^2 Y^{(s)}, \quad
       Y_\alpha^{(s)} \equiv - {1 \over k} Y_{,\alpha}^{(s)}, 
   \nonumber \\
   & & Y_{\alpha\beta}^{(s)} \equiv {1 \over k^2} Y_{,\alpha|\beta}^{(s)}
       + {1 \over 3} g^{(3)}_{\alpha\beta} Y^{(s)},
   \nonumber \\
   & & Y^{(v)|\gamma}_{\alpha\;\;\;\;\;\;\;\gamma}
       \equiv - k^2 Y^{(v)}_\alpha, \quad
       Y_{\alpha\beta}^{(v)}
       \equiv - {1 \over k} Y_{(\alpha|\beta)}^{(v)}, \quad
       Y_\alpha^{(v)|\alpha} \equiv 0,
   \nonumber \\
   & & Y^{(t)|\gamma}_{\alpha\beta\;\;\;\;\gamma}
       \equiv - k^2 Y^{(t)}_{\alpha\beta}, \quad
       Y^{(t)}_{\alpha\beta} \equiv Y^{(t)}_{\beta\alpha}, \quad
       Y^{(t)\alpha}_{\;\;\;\;\;\alpha} \equiv 0 
       \equiv Y_{\alpha\beta}^{(t)|\beta},
   \nonumber \\
   \label{Y-def}
\eea
where ${\bf k}$ is a wave vector in Fourier space with $k = |{\bf k}|$;
the wave vector for individual type of perturbation is defined by
the Helmholtz equations in eq. (\ref{Y-def}).
In terms of the harmonic functions we have
$\alpha ({\bf x}, \eta) \equiv \alpha ({\bf k}, \eta) 
Y^{(s)} ({\bf k}; {\bf x})$ and similarly for $\beta$, $\gamma$ and $\varphi$;
$B^{(v)}_\alpha \equiv b^{(v)} Y_\alpha^{(v)}$,
$C^{(v)}_\alpha \equiv c^{(v)} Y_\alpha^{(v)}$, and
$C^{(t)}_{\alpha\beta} \equiv c^{(t)} Y^{(t)}_{\alpha\beta}$.
Since we are considering the linear perturbations the same forms of
equations will be valid in the configuration and the Fourier spaces. 
Thus, without causing any confusion, we often ignore distinguishing the 
Fourier space from the configuration space by an additional subindex.
Also, since each Fourier mode evolves independently to the linear order,
without causing any confusion we ignore the summation over eigenfunctions
indicating the Fourier expansion.

The perturbed scalar fields  $\delta \phi^I$ in eq. (\ref{phi-pert}) 
only couple with the scalar-type perturbations, and are expanded as 
\bea
   & & \delta \phi^I ({\bf x}, t) = \delta \phi^I ({\bf k},t) 
       Y^{(s)} ({\bf k}; {\bf x}),
\eea
and are similarly for $\delta R$ and $\delta F$ as well.

Now, we consider perturbations in the fluid quantities.
We decompose $v_\alpha$ and $\pi^{(3)\alpha}_{\;\;\;\;\;\beta}$
into three-types of perturbations as
\bea
   & & v_\alpha \equiv v^{(s)} Y_\alpha^{(s)} + v^{(v)} Y_\alpha^{(v)},
   \nonumber \\
   & & \pi_{\;\;\;\;\;\alpha}^{(3)\beta} \equiv 
       \pi^{(s)} Y^{(s)\beta}_{\;\;\;\;\;\alpha}
       + \pi^{(v)} Y^{(v)\beta}_{\;\;\;\;\;\alpha}
       + \pi^{(t)} Y^{(t)\beta}_{\;\;\;\;\;\alpha}.
   \label{fluid-decomp}
\eea
The energy-momentum tensor in eq. (\ref{Tab-decomp}) becomes
\bea
   & & T^0_0 = - \left( \bar \mu + \delta \mu \right),
   \nonumber \\
   & & T^0_\alpha = - {1 \over k} \left( \mu + p \right) v^{(s)}_{,\alpha}
       + \left( \mu + p \right) v^{(v)} Y^{(v)}_{\alpha},
   \nonumber \\
   & & T^\alpha_\beta = \left( \bar p + \delta p \right) \delta^\alpha_\beta
       + \pi^{(s)} Y^{(s)\alpha}_{\;\;\;\;\;\beta}
       + \pi^{(v)} Y^{(v)\alpha}_{\;\;\;\;\;\beta}
       + \pi^{(t)} Y^{(t)\alpha}_{\;\;\;\;\;\beta}.
   \nonumber \\
   \label{Tab}
\eea
In terms of the individual matter's fluid quantities we have
\bea
   & & \bar \mu = \sum_{l} \bar \mu_{(l)}, \quad
       \delta \mu = \sum_{l} \delta \mu_{(l)}, 
   \label{fluid-sum} 
\eea
and similarly for $\bar p$, $\delta p$, $(\mu + p) v^{(s,v)}$, and
$\pi^{(s,v,t)}$.
We use the notation introduced by Bardeen in 1988 \cite{Bardeen-1988};
comparison with Bardeen's 1980 notation \cite{Bardeen-1980} can be found
in \S 2.2 of \cite{Hwang-1991-PRW};
compared with our previous notation in \cite{Hwang-1991-PRW} we have
$\pi^{(s)} = (k^2/a^2) \sigma$ and $v^{(s)} = - k/[a(\mu+p)] \Psi$.
We often write $v \equiv v^{(s)}$.

The interaction terms among fluids introduced in 
eq. (\ref{Tab-i-conservation}) are decomposed as:
\bea
   & & Q_{(i)0} \equiv - a \left[ \bar Q_{(i)} ( 1 + A ) 
       + \delta Q_{(i)} \right],
   \nonumber \\
   & & Q_{(i)\alpha} \equiv J^{(s)}_{(i)} Y^{(s)}_{,\alpha}
       + J^{(v)}_{(i)} Y^{(v)}_{\alpha}.
\eea
{}From eq. (\ref{Tab-i-conservation}) we have
\bea
   & & \sum_{l} \bar Q_{(l)} = \bar \Gamma_I \dot {\bar \phi}^I, \quad
   \nonumber \\
   & & \sum_{l} \delta Q_{(l)} = \delta \Gamma_I \dot \phi^I 
       + \Gamma_I ( \delta \dot \phi - \dot \phi \alpha ),
   \nonumber \\
   & & \sum_{l} J^{(s)}_{(l)} = - \Gamma_I \delta \phi^I, \quad
       \sum_{l} J^{(v)}_{(l)} = 0.
   \label{Q-sum}
\eea
Thus, RHS of the second equation in eq. (\ref{Tab-i-conservation}) 
contributes only to the scalar-type perturbation.

\subsection{Background equations}
                                          \label{sec:background}

The equations for the background are: 
\bea
   & & H^2 = {1 \over 3F} \left[ \mu
       + {1 \over 2} g_{IJ} \dot \phi^I \dot \phi^J
       - {1 \over 2} \left( f - RF - 2V \right) - 3 H \dot F \right]
   \nonumber \\
   & & \quad
       - {K \over a^2},
   \label{BG1} \\
   & & \dot H = - {1\over 2 F} \left( \mu + p + g_{IJ} \dot \phi^I \dot \phi^J
       + \ddot F - H \dot F \right) + {K \over a^2},
   \label{BG2} \\
   & & R = 6 \left( 2 H^2 + \dot H + {K \over a^2} \right),
   \label{BG3} \\
   & & \ddot \phi^I + 3 H \dot \phi^I + \Gamma^I_{JK} \dot \phi^J \dot \phi^K
       + {1 \over 2} \left( 2 V - f \right)^{;I} = - \Gamma^I,
   \label{BG4} \\
   & & \dot \mu_{(i)} + 3 H \left( \mu_{(i)} + p_{(i)} \right) = Q_{(i)},
   \label{BG5}
\eea
where $\mu$, $p$ and $Q_{(i)}$ follow eqs. (\ref{fluid-sum},\ref{Q-sum}).
Equations (\ref{BG1},\ref{BG2}) follow from $G^0_0$ and 
$G^\alpha_\alpha - 3 G^0_0$ components of Eq. (\ref{GFE}), respectively.
Equation (\ref{BG4}) follows from eq. (\ref{EOM}).
Equation (\ref{BG5}) follows from eq. (\ref{Tab-i-conservation}).
By adding eq. (\ref{BG5}) over components we have
\bea
   & & \dot \mu + 3 H ( \mu + p ) = \Gamma_I \dot \phi^I.
   \label{BG6}
\eea

By setting $F = 1/(8 \pi G)$ we can recover $8 \pi G$ factor in 
Einstein gravity. 
The gravity theory in eq. (\ref{Lagrangian}) includes the cosmological 
constant, $\Lambda$.
The cosmological constant introduced in eq. (\ref{Lagrangian}) as an
additional $-\Lambda$ term can be simulated using either the scalar field
or the fluid. 
Using the scalar field we let $V \rightarrow V + \Lambda/(8 \pi G)$.
Using the fluid, since $\Lambda$ contributes 
$T_{ab}^{\Lambda} = - \Lambda g_{ab}/(8 \pi G)$ to the energy-momentum tensor, 
we let $\mu \rightarrow \mu + \Lambda/(8 \pi G)$ and 
$p \rightarrow p - \Lambda/(8 \pi G)$.
This causes a change only in eq. (\ref{BG1}).
In the presence of the kinetic components we additionally have
the Boltzmann equations for the components and the sum over fluid
quantities should include the contributions from the kinetic
components, see \S \ref{sec:BG-evol}.

\subsection{Gauge strategy}
                                       \label{sec:gauge}

In the following we explain briefly our gauge-ready strategy.
Due to the general covariance of the relativistic gravity theory
we need to take care of the fictitious degrees of freedom arising in the
relativistic perturbation analysis.
This freedom appears because the relativistic gravity is a constrained 
system: there exist some constraint equations with only algebraic relations
among variables.
In the perturbation analysis this is known as the gauge degree of freedom.
The gauge freedom in the perturbation analysis arises from different ways 
of defining the correspondence between the perturbed spacetime and the 
fictitious background.
{}For example, by introducing a spacetime dependent coordinate
transformation, even the FLRW background can be changed into
a perturbed form which is simply due to the coordinate (gauge) transformation.
Only in a special coordinate system the FLRW metric looks simple as in
eq. (\ref{metric-general}) without perturbations.

Similarly as in other gauge theories, there are some redundant degrees of 
freedom in the equations which can be fixed without affecting the physics.
Certainly it would be advisable, and is often essential, to take a proper 
gauge condition which either simplifies the mathematical analyses or 
allows an easier physical interpretation.
Usually we do not know the best gauge condition (which differs depending 
on each problem) {\it a priori}, but it is desirable (actually often 
necessary) to find out the best one.
In this regards, the advantage of managing the equations in a gauge-ready 
form was suggested by Bardeen in 1988 \cite{Bardeen-1988}, 
and the formulation was elaborated in \cite{Hwang-1991-PRW}.

Contrary to many works in the literature which often refers the gauge 
freedom causing problems in the theory, we believe that, as in the other 
gauge theories (e.g., the Maxwell theory and the Yang-Mills theory), 
the gauge freedom can/should be used as an {\it advantage} over 
solving each specific problem.
Our set of gauge-ready form arrangement of the equations,
by allowing simple adoptation of different gauge conditions, will allow
the optimal use of the advantageous aspect of the gauge degrees of freedom
present in the theory.
To that purpose all the scalar-type perturbation equations are presented
in a uniquely significant (see below) spatially gauge-invariant form 
but without fixing the temporal gauge condition.
In this way, we can easily implement the several available temporal 
gauge conditions depending on the situation, and
in this sense, the set of equations is in a gauge-ready form.
The tensor-type perturbation describing the gravitational wave is
gauge-invariant, and the vector-type perturbation describing the rotation 
is presented using the uniquely significant gauge-invariant combinations 
of the variables.
The particular choice of a gauge implies no loss of generality.
If a solution of a variable is known in a specific gauge, the rest of the 
variables, even in the other gauges, can be easily recovered.
Therefore, if possible, it would be convenient to start from the 
gauge condition which allows an easier manipulation of the equations.
However, since the optimum gauge condition is usually unknown {\it a priori},
often it is convenient to carry out the analyses in the available
pool of various gauge conditions and to find out the distinguished
gauge condition; such analyses in the single component situations were 
carried out in the fluid \cite{Hwang-Ideal-1993}, in the scalar field 
\cite{Hwang-MSF-1994}, and in the generalized gravity theories 
\cite{Hwang-Noh-1996-GGT}.
Our experience tells that different gauge conditions fit different problems,
or even different aspects of a given system. 
Often, problematic aspects of the gauge freedom appear if one sticks
to a particular gauge condition from the beginning and if that gauge 
condition turns out to be not a suitable choice for the problem.
Our gauge-ready strategy is not 
a particularly new suggestion in the contex of the gauge theory
except that such a strategy, and its {\it systematic use}, has been largely 
ignored in the cosmology literature despite its rather apparent advantage.
In the present work we extend the formulation in \cite{Hwang-1991-PRW} 
to more general situation including kinetic components and arrange the 
equations for the convenient usage in diverse situations.

In the perturbation analyses we have to deal with two metric systems,
one is the physical perturbed model and the other is the fictitious
background model.
The gauge degrees of freedom arise because we have different ways of
corresponding the perturbed spacetime points with the arbitrary
background spacetime points.
Since we are considering the spatially homogeneous and isotropic background 
the spatial correspondences (spatial gauge transformation) can be handled 
trivially: according to Bardeen \cite{Bardeen-1988} 
``Since the background 3-space is homogeneous
and isotropic, the perturbation in all physical quantities must in fact
be gauge invariant under purely spatial gauge transformations.''
We will show that only the variables $\beta$, $\gamma$, $b^{(v)}$,
and $c^{(v)}$ depend on the spatial gauge transformation. 
But these appear always in the combinations $\chi$ and $\Psi^{(v)}$
in eq. (\ref{chi-kappa}) which are spatially gauge-invariant combinations;
see eq. (\ref{GT-2}).
These combinations are unique in the sense that other combinations fail 
to fix the spatial gauge degrees of freedom completely.
Thus, using these (uniquely significant) spatially gauge-invariant 
combinations we take care of the effects of spatial gauge 
transformation of the scalar- and vector-type perturbations completely;
the corresponding spatial gauge transformation properties of the
kinetic components will be considered below eq. (\ref{GT-f}).

Gauge transformation properties of the perturbed cosmological spacetime
were nicely discussed in 
\cite{Stewart-Walker-1974,%
Bardeen-1980,%
Kodama-Sasaki-1984,%
Bardeen-1988,%
Ellis-Bruni-1989}.
Under the gauge transformation of the form $\tilde x^a = x^a + \xi^a$
the metric and the energy-momentum tensor transform as
\bea
   & & \tilde g_{ab} (\tilde x^e) = {\partial x^c \over \partial \tilde x^a}
       {\partial x^d \over \partial \tilde x^b} g_{cd} (x^e), 
\eea
thus,
\bea
   & & \tilde g_{ab} (x^e) = g_{ab} (x^e)
       - g_{ab,c} \xi^c - g_{bc} \xi^c_{\;\;,a} - g_{ac} \xi^c_{\;\;,b},
   \label{gab-GT}
\eea
and similarly for $T^a_b$.
By introducing $\xi^t \equiv a \xi^0$ ($0 = \eta$) and 
$\xi_\alpha \equiv \xi_{,\alpha} + \xi_\alpha^{(v)}$, with
$\xi_\alpha^{(v)}$ based on $g^{(3)}_{\alpha\beta}$ and
$\xi^{(v)|\alpha}_\alpha = 0$, the perturbed metric quantities 
and the collective fluid quantities change as:
\bea
   & & \tilde \alpha = \alpha - \dot \xi^t, \quad
       \tilde \varphi = \varphi - H \xi^t, \quad
       \tilde \beta = \beta - {1 \over a} \xi^t + a \dot \xi, 
   \nonumber \\
   & & \tilde \gamma = \gamma - \xi, \quad
       \delta \tilde \mu = \delta \mu - \dot \mu \xi^t, \quad
       \delta \tilde p = \delta p - \dot p \xi^t, 
   \nonumber \\
   & & \tilde v = v - {k \over a} \xi^t, 
   \nonumber \\
   & & \tilde B_\alpha^{(v)} = B_\alpha^{(v)} + a \dot \xi_\alpha^{(v)}, \quad
       \tilde C_\alpha^{(v)} = C_\alpha^{(v)} - \xi_\alpha^{(v)}, 
   \label{GT-1}
\eea
and $v^{(v)}$, $C^{(t)}_{\alpha\beta}$, $\pi^{(s,v,t)}$ are gauge invariant.
Thus, from eq. (\ref{chi-kappa}) we have
\bea
   & & \tilde \chi = \chi - \xi^t, \quad
       \tilde \kappa = \kappa + \left( 3 \dot H + {\Delta \over a^2}
       \right) \xi^t, 
   \nonumber \\
   & & \tilde \Psi^{(v)} = \Psi^{(v)},
   \label{GT-2}
\eea
and these are spatially gauge-invariant.
{}From the scalar nature of $\phi^I$, $R$, $F$, and $\Gamma_I$ we have:
\bea
   & & \delta \tilde \phi^I = \delta \phi^I - \dot \phi^I \xi^t, \quad
       \delta \tilde \Gamma_I = \delta \Gamma_I - \dot \Gamma_I \xi^t, 
   \nonumber \\
   & & \delta \tilde R = \delta R - \dot R \xi^t, \quad
       \delta \tilde F = \delta F - \dot F \xi^t.
   \label{GT-3}
\eea

{}From eq. (\ref{GT-1}) we notice that the tensor-type perturbation
variables are gauge-invariant.
{}For the vector-type perturbation we notice that $\Psi^{(v)}$ 
defined in eq. (\ref{chi-kappa}) is a unique gauge-invariant combination.
Thus, using $\Psi^{(v)}$ the vector-type perturbation becomes gauge-invariant.
{}For the scalar-type perturbation using $\chi$ instead of $\beta$
and $\gamma$ individually, all the variables are spatially gauge invariant.
Considering the temporal gauge transformation properties
we notice several fundamental 
gauge conditions based on the metric and the energy-momentum tensor:
\begin{tabbing}
\hskip 1cm \= Synchronous gauge: \hskip 2cm \= $\alpha \equiv 0$,      \\
           \> Comoving gauge:               \> $v/k \equiv 0$,         \\
           \> Zero-shear gauge:             \> $\chi \equiv 0$,        \\
           \> Uniform-curvature gauge:      \> $\varphi \equiv 0$,     \\
           \> Uniform-expansion gauge:      \> $\kappa \equiv 0$,      \\
           \> Uniform-density gauge:        \> $\delta \mu \equiv 0$,  \\
           \> Uniform-pressure gauge:       \> $\delta p \equiv 0$,    \\
           \> Uniform-field ($\phi^I$) gauge:  \> $\delta \phi^I \equiv 0$, \\
           \> Uniform-$R$ gauge:            \> $\delta R \equiv 0$,    \\
           \> Uniform-$F$ gauge:            \> $\delta F \equiv 0$,  
\end{tabbing}
\bea
   \label{Gauges-1}
\eea
etc.
The names of the gauge conditions using $\chi$, $\varphi$ and $\kappa$
can be justified: these variables correspond to the shear, 
the three-space curvature and the perturbed expansion
of the normal-frame vector field, respectively, 
see eqs. (\ref{kinematic-n},\ref{ADM-variables}). 
$v/k \equiv 0$ is a frame-invariant definition of the comoving gauge 
condition based on the collective velocity.

The original definition of the synchronous gauge in \cite{Lifshitz-1946}
fixed $\beta = 0$ as the spatial gauge condition in addition to 
$\alpha = 0$ as the temporal gauge condition.
In such a case, from eq. (\ref{GT-1}) we notice that the spatial
gauge fixing also leaves remaining (spatial) gauge degree of freedom.
By using the spatially gauge-invariant combinations
$\chi$ and $v$ we can avoid such an unnecessary complication
caused by the spatial gauge transformation
which is trivial due to the homogeneity of the FLRW background
\cite{Bardeen-1988}.
{}From eq. (\ref{chi-kappa}) $\chi$ is the same as $a \beta$ in the
$\gamma = 0$ gauge condition. 
But in the $\beta = 0$ gauge condition we have $\chi = a^2 \dot \gamma$, 
thus $\gamma$ is undetermined up to a constant (in time only)
factor which is the (spatially varying) remaining gauge mode.

By examining eqs. (\ref{GT-1},\ref{GT-2},\ref{GT-3},\ref{Gauges-1}) 
we notice that,
out of the several gauge conditions in eq. (\ref{Gauges-1}),
except for the synchronous gauge condition, each of the gauge conditions
fixes the temporal gauge mode completely; the synchronous gauge,
$\alpha = 0$, leaves spatially varying nonvanishing $\xi^t ({\bf x})$ 
which is the remaining gauge mode even after the gauge fixing.
Thus, a variable in such a gauge condition uniquely corresponds
to a gauge-invariant combination which combines the variable concerned
and the variable used in the gauge condition.
Several interesting gauge-invariant combinations are the following:
\bea
   & & \delta \mu_{v} \equiv \delta \mu - {a \over k} \dot \mu v, 
       \quad \varphi_\chi \equiv \varphi - H \chi, \quad
       v_\chi \equiv v - {k \over a} \chi, 
   \nonumber \\
   & & \varphi_v \equiv \varphi - {aH \over k} v, \quad
       \delta \phi^I_\varphi \equiv \delta \phi^I 
       - {\dot \phi^I \over H} \varphi
       \equiv - {\dot \phi \over H} \varphi_{\delta \phi^I}.
   \label{GI-single}
\eea
{}For example, the gauge-invariant combination $\delta \phi^I_\varphi$ 
is equivalent to $\delta \phi^I$ in the uniform-curvature gauge which takes 
$\varphi \equiv 0$ as the gauge condition, etc.
In this way, we can systematically construct various gauge-invariant
combinations for a given variable.
Since we can make several gauge-invariant combinations even for a given
variable, this way of writing the gauge-invariant combination will turn out 
to be convenient practically.

In the multi-component case of fluids there exist some 
additional (temporal) gauge conditions available.
{}From the tensorial property of $T_{(i)ab}$ and using eq. (\ref{gab-GT})
we can show
\bea
   & & \delta \tilde \mu_{(i)} = \delta \mu_{(i)} - \dot \mu_{(i)} \xi^t, \quad
       \delta \tilde p_{(i)} = \delta p_{(i)} - \dot p_{(i)} \xi^t, 
   \nonumber \\
   & & \tilde v_{(i)} = v_{(i)} - {k \over a} \xi^t,
   \label{GT-i-1}
\eea
and $v^{(v)}_{(i)}$, $\pi^{(s,v,t)}_{(i)}$ are gauge invariant.
Thus, the additional temporal gauge conditions are:
\bea
   & & \delta \mu_{(i)} \equiv 0, \quad
       \delta p_{(i)} \equiv 0, \quad
       v_{(i)}/k \equiv 0, \quad
       \delta \phi^I \equiv 0, \quad
   \label{Gauges-2}
\eea
etc.
Any one of these gauge conditions also fixes the temporal gauge condition
completely.
{}From the vector nature of $Q_{(i)a}$ and using 
eq. (\ref{Tab-i-conservation}) we have
\bea
   & & \delta \tilde Q_{(i)} = \delta Q_{(i)} - \dot Q_{(i)} \xi^t, \quad
       \tilde J^{(s)}_{(i)} = J^{(s)}_{(i)} + Q_{(i)} \xi^t, 
   \nonumber \\
   & & \tilde J^{(v)}_{(i)} = J^{(v)}_{(i)}.
   \label{GT-i-3}
\eea

As mentioned previously, in general we do not know the suitable gauge
condition {\it a priori}.
The proposal made in \cite{Bardeen-1988,Hwang-1991-PRW} is that we write 
the set of equation without fixing the (temporal) gauge condition 
and arrange the equation so that we can implement easily various 
fundamental gauge conditions.
We call this approach a {\it gauge-ready method}.
Any one of the fundamental gauge conditions in 
eqs. (\ref{Gauges-1},\ref{Gauges-2}) and {\it suitable linear 
combinations} of them can turn out to be a useful gauge condition 
depending on the problem.
A particular gauge condition is 
suitable for handling a particular aspect of the individual problem.
The gauge transformation properties of the kinetic components will be
considered in \S \ref{sec:Kinetic};
see paragraphs surrounding eqs. (\ref{GT-f},\ref{GT-Theta},\ref{GT-hat-Theta}).

\subsection{Scalar-type perturbation}
                                          \label{sec:scalar}

In this section we present a complete set of equations describing 
the scalar-type perturbation without fixing the temporal gauge condition,
i.e., in the gauge-ready form.

\noindent
Definition of $\kappa$:
\bea
   & & \dot \varphi = H \alpha - {1\over 3} \kappa
       + {1\over 3} {k^2 \over a^2} \chi.
   \label{G1} 
\eea
ADM energy constraint ($G^0_0$ component of the field equation):
\bea
   & & - {k^2 - 3K \over a^2} \varphi
       + \left( H + {\dot F \over 2 F} \right) \kappa
       - {1\over 2F} \left( g_{IJ} \dot \phi^I \dot \phi^J
       - 3 H \dot F \right) \alpha
   \nonumber \\
   & & \quad
       = - {1\over 2F} \Bigg\{ \delta \mu
       + g_{IJ} \dot \phi^I \delta \dot \phi^J
   \nonumber \\
   & & \quad
       + {1 \over 2} \left[ g_{IJ,K} \dot \phi^I \dot \phi^J
       - \left( f - 2 V \right)_{,K} \right] \delta \phi^K
   \nonumber \\
   & & \quad
       - 3 H \delta \dot F
       + \left( 3 \dot H + 3 H^2 - {k^2 \over a^2} \right) \delta F \Bigg\}.
   \label{G2}
\eea
Momentum constraint ($G^0_\alpha$ component):
\bea
   & & \kappa - {k^2 - 3 K \over a^2} \chi
       + {3\over 2} {\dot F \over F} \alpha
       = {3 \over 2F} \Big[ {a \over k} \left( \mu + p \right) v
       + g_{IJ} \dot \phi^I \delta \phi^J 
   \nonumber \\
   & & \quad
       + \delta \dot F - H \delta F \Big].
   \label{G3}
\eea
ADM propagation 
($G^\alpha_\beta - {1 \over 3} \delta^\alpha_\beta G^\gamma_\gamma$ component):
\bea
   & & \dot \chi + \left( H + {\dot F \over F} \right) \chi
       - \alpha - \varphi = {1 \over F} \left( {a^2 \over k^2} \pi^{(s)}
       + \delta F \right).
   \label{G4}
\eea
Raychaudhuri equation ($G^\gamma_\gamma - G^0_0$ component):
\bea
   & & \dot \kappa 
       + \left( 2 H + {\dot F \over 2F} \right) \kappa
       + {3 \over 2} {\dot F \over F} \dot \alpha
   \nonumber \\
   & & \quad
       + \left[ 3 \dot H + {1 \over 2 F} \left( 6 \ddot F + 3 H \dot F
       + 4 g_{IJ} \dot \phi^I \dot \phi^J \right)
       - {k^2 \over a^2} \right] \alpha
   \nonumber \\
   & & \quad
       = {1 \over 2F} \Bigg\{ \delta \mu + 3 \delta p
       + 4 g_{IJ} \dot \phi^I \delta \dot \phi^J
   \nonumber \\
   & & \quad
       + \left[ 2 g_{IJ,K} \dot \phi^I \dot \phi^J
       + \left( f - 2 V \right)_{,K} \right] \delta \phi^K
   \nonumber \\
   & & \quad
       + 3 \delta \ddot F + 3 H \delta \dot F
       + \left( - 6 H^2 + {k^2 - 6 K \over a^2} \right) \delta F \Bigg\}.
   \label{G5}
\eea
Scalar fields equations of motion:
\bea
   & & \delta \ddot \phi^I + 3 H \delta \dot \phi^I
       + 2 \Gamma^I_{JK} \dot \phi^J \delta \dot \phi^K
       + {k^2 \over a^2} \delta \phi^I
   \nonumber \\
   & & \quad
       + \left[ {1 \over 2} \left( 2 V - f \right)^{;I}_{\;\;\;, L}
       + \Gamma^I_{JK,L} \dot \phi^J \dot \phi^K \right] \delta \phi^L
   \nonumber \\
   & & \quad
       = \dot \phi^I \left( \kappa + \dot \alpha \right)
       + \left( 2 \ddot \phi^I + 3 H \dot \phi^I
       + 2 \Gamma^I_{JK} \dot \phi^J \dot \phi^K \right) \alpha
   \nonumber \\
   & & \quad
       + {1 \over 2} F^{;I} \delta R - \delta \Gamma^I.
   \label{G6}
\eea
Trace equation ($G^a_a$ component):
\bea
   & & \delta \ddot F + 3 H \delta \dot F + \left( {k^2 \over a^2}
       - {R \over 3} \right) \delta F
       + {2\over 3} g_{IJ} \dot \phi^I \delta \dot \phi^J
   \nonumber \\
   & & \quad
       + {1 \over 3} \left[ g_{IJ,K} \dot \phi^I \dot \phi^J
       + 2 \left( f - 2 V \right)_{,K} \right] \delta \phi^K
   \nonumber \\
   & & \quad
       = {1 \over 3} \left( \delta \mu - 3 \delta p \right)
       + \dot F \left( \kappa + \dot \alpha \right)
   \nonumber \\
   & & \quad
       + \left( {2 \over 3} g_{IJ} \dot \phi^I \dot \phi^J
       + 2 \ddot F + 3 H \dot F \right) \alpha
       - {1\over 3} F \delta R.
   \label{G7}
\eea
Scalar curvature:
\bea
   & & \delta R = 2 \Big[ - \dot \kappa - 4 H \kappa
       + \left( {k^2 \over a^2} - 3 \dot H \right) \alpha
   \nonumber \\
   & & \quad
       + 2 {k^2 - 3 K \over a^2} \varphi \Big].
   \label{G8}
\eea
Energy conservation of the fluid components 
[from $T_{(i)0;b}^{\;\;\;\; b} = Q_{(i)0}$ and using eq. (\ref{G1})]:
\bea
   & & \delta \dot \mu_{(i)} + 3 H \left( \delta \mu_{(i)} 
       + \delta p_{(i)} \right)
       = - {k \over a} \left( \mu_{(i)} + p_{(i)} \right) v_{(i)} 
       + \dot \mu_{(i)} \alpha
   \nonumber \\
   & & \quad
       + \left( \mu_{(i)} + p_{(i)} \right) \kappa + \delta Q_{(i)}.
   \label{G9}
\eea
Momentum conservation of the fluid components 
(from $T_{(i)\alpha;b}^{\;\;\;\; b} = Q_{(i)\alpha}$):
\bea
   & & {1 \over a^4 ( \mu_{(i)} + p_{(i)} )}
       \left[ a^4 ( \mu_{(i)} + p_{(i)} ) v_{(i)} \right]^\cdot
       = {k \over a} \Bigg[ \alpha 
   \nonumber \\
   & & \quad
       + {1 \over \mu_{(i)} + p_{(i)}} \left( \delta p_{(i)} 
       - {2 \over 3} {k^2 - 3 K \over k^2} \pi^{(s)}_{(i)} - {J}_{(i)} \right)
       \Bigg].
   \label{G10}
\eea
By adding properly eqs. (\ref{G9},\ref{G10}) over all components of the fluids,
and using properties in eqs. (\ref{fluid-sum},\ref{Q-sum},\ref{BG6}),
we get equations for the collective fluid quantities as:
\bea
   & & \delta \dot \mu + 3 H \left( \delta \mu + \delta p \right)
       = ( \mu + p ) \left( \kappa - 3 H \alpha - {k \over a} v \right)
   \nonumber \\
   & & \quad
       + \Gamma_I \delta \dot \phi^I + \delta \Gamma_I \dot \phi^I,
   \label{G11} \\
   & & {1 \over a^4 ( \mu + p )} \left[ a^4 ( \mu + p ) v \right]^\cdot
   \nonumber \\
   & & \quad
       = {k \over a} \Bigg[ \alpha + {1 \over \mu + p} \left( \delta p 
       - {2 \over 3} {k^2 - 3 K \over k^2} \pi^{(s)} 
       + \Gamma_I \delta \phi^I \right) \Bigg].
   \nonumber \\
   \label{G12}
\eea
It is convenient to introduce
\bea
   & & \delta p (k, t) \equiv c_s^2 (t) \delta \mu (k, t)
       + e (k, t), \quad
       \delta \equiv {\delta \mu \over \mu}, 
   \nonumber \\
   & & w (t) \equiv {p \over \mu}, \quad
       c_s^2 (t) \equiv {\dot p \over \dot \mu}.
   \label{w-cs}
\eea

Equations (\ref{G1}-\ref{G12}) provide a redundantly complete set for handling 
the most general scalar-type perturbation of the FLRW world model
allowed by the Lagrangian in eq. (\ref{Lagrangian});
for example, eq. (\ref{G7}) follows from eqs. (\ref{G2},\ref{G5},\ref{G8}).
Equations (\ref{G11},\ref{G12}) follow from eqs. (\ref{G9},\ref{G10}).
Following the prescriptions below eq. (\ref{BG6})
these equations also include the cosmological constant;
an introduction of $\Lambda$ term does not appear explicitly in
our set of equations in the form eqs. (\ref{G1}-\ref{G12}).
Notice that eqs. (\ref{G9}-\ref{G12}), which follow from fluid 
energy-momentum conservation in eqs. (\ref{Tab-f},\ref{Tab-i-conservation}), 
are not affected formally by the generalized nature of the gravity 
we are considering.
In \S \ref{sec:Kinetic} and \ref{sec:CMB}
we will see that the presence of kinetic components
additionally introduces the corresponding Boltzmann equations,
and their contributions to the energy-momentum content can be included
as the individual fluid quantity in the above set of equations.

Equations (\ref{G1}-\ref{G12}) are written in a gauge-ready form.
In handling the actual problem we have a {\it right} to impose {\it one} 
temporal gauge condition according to the mathematical or physical 
conveniences we can achieve.
As long as we choose a gauge condition which fixes the temporal gauge
mode completely, the resulting equations and the solutions are
completely free from the gauge degree of freedoms and variables are 
equivalently gauge-invariant.
Some recommended fundamental gauge conditions are summarized in 
eqs. (\ref{Gauges-1},\ref{Gauges-2}).
Equations (\ref{G1}-\ref{G12}) are designed so that we can easily
accomodate any of these gauge conditions.

If we take an {\it ansatz} for $\Gamma^I$ term in eq. (\ref{EOM}) as
\bea
   & & \Gamma^I \equiv D^I_J \phi^J_{\;\;;a} u^a,
\eea
to the perturbed order we have
\bea
   & & \bar \Gamma^I + \delta \Gamma^I \equiv \bar D^I_J \dot {\bar \phi}^J
       + D^I_J \delta \dot \phi^J - D^I_J \dot \phi^J \alpha
       + \delta D^I_J \dot \phi^J,
   \nonumber \\
\eea
where we used $u^0 = {1 \over a} ( 1- \alpha)$.
Such a phenomenological damping term was considered in
\cite{Salopek-etal-1989}.

\subsection{Rotation}
                                       \label{sec:rotation}

The equations for the vector-type (rotational) perturbation are:
\bea
   & & {k^2 - 2 K \over 2 a^2} \Psi^{(v)}
       = {1 \over F} \sum_{l} \left( \mu_{(l)} + p_{(l)} \right) 
       v^{(v)}_{(l)},
   \label{rot-1} \\
   & & {1 \over a^3} \left[ a^4 \left( \mu + p \right) v^{(v)}
       \right]^\cdot = - {k^2 - 2 K \over 2k} \pi^{(v)},
   \label{rot-2} \\
   & & {1 \over a^3} \left[ a^4 \left( \mu_{(i)} + p_{(i)} \right)
       v^{(v)}_{(i)} \right]^\cdot
       = - {k^2 - 2 K \over 2k} \pi^{(v)}_{(i)} + J^{(v)}_{(i)}.
   \nonumber \\
   \label{rot-i}
\eea
Equations (\ref{rot-1}) follows from $G^0_\alpha$ component of Eq. (\ref{GFE}), 
and eq. (\ref{rot-2}) follows from $T^b_{\alpha ; b} = 0$.
Equation (\ref{rot-i}) follows from eq. (\ref{Tab-i-conservation}).
By adding eq. (\ref{rot-i}) over all components we have eq. (\ref{rot-2}).
Notice that eqs. (\ref{rot-2},\ref{rot-i}) 
are not affected formally by the generalized nature of gravity theory.
In fact, these two equations are derived from the conservations of
the energy-momentum tensors in 
eqs. (\ref{fluid-conservation},\ref{Tab-i-conservation})
{\it without using} the gravitational field equation.
The presence of kinetic components
additionally introduces the corresponding Boltzmann equations,
and contributes to the fluid quantities in the above equations,
see \S \ref{sec:Kinetic} and \ref{sec:CMB}.

The vorticity tensors based on the frame-invariant four-vectors are
(see the Appendix \ref{sec:Kinematic}):
\bea
   & & \omega_{\alpha\beta} = a v^{(v)} Y^{(v)}_{[\alpha|\beta]}, \quad
       \omega_{(i)\alpha\beta}
       = a v^{(v)}_{(i)} Y^{(v)}_{[\alpha|\beta]}.
\eea
Thus, we have
$\omega \equiv \sqrt{ \omega^{ab} \omega_{ab}/2 }$
and similarly for $\omega_{(i)}$.
Equations (\ref{rot-1}-\ref{rot-i}) show that the fluid velocities of the 
rotational perturbation do not explicitly depend on the generalized nature 
of the gravity, whereas, only the metric connected with the rotation mode, 
$\Psi^{(v)}$, depends on the nature of generalized gravity;
$\Psi^{(v)}$ term appears in the Boltzmann equations though, see
eqs. (\ref{M-matrix},\ref{hat-M-matrix}).
Equations (\ref{rot-2},\ref{rot-i}) which are independent of the 
field equations tell that in a medium without anisotropic
stress terms $\pi^{(v)}_{(i)}$ 
and the mutual interaction terms among components ${J}_{(i)}$,
the angular momentum combination of individual component is conserved as
\bea
   {\rm Angular \; Momentum} 
   &\sim& a^3 \left( \mu_{(i)} + p_{(i)} \right) 
       \times a \times v^{(v)}_{(i)} 
   \nonumber \\
   &=& {\rm constant \; in \; time}.
\eea
The presence of anisotropic pressure can work as the sink or the source of
the rotational perturbation of the individual fluid.
Conservation of the angular momentum combination of the rotational 
perturbation in Einstein gravity was noticed in the original work 
by Lifshitz \cite{Lifshitz-1946}.

In the presence of kinetic components we additionally have the 
corresponding Boltzmann equations, and the components contribute 
to the anisotropic pressure in the above equations,
see \S \ref{sec:Kinetic} and \ref{sec:CMB}.

\subsection{Gravitational wave}
                                       \label{sec:GW}

The tensor-type perturbation (gravitational wave) equation in Einstein gravity 
was derived originally by Lifshitz in \cite{Lifshitz-1946}.
We can derive easily the wave equation for the most general situation 
covered by the Lagrangian in eq. (\ref{Lagrangian}) as
\bea
   & & \ddot c^{(t)} + \left( 3 H + {\dot F \over F} \right) \dot c^{(t)}
       + {k^2 + 2 K \over a^2} c^{(t)}
       = {1 \over F} \sum_{l} \pi^{(t)}_{(l)},
   \nonumber \\
   \label{GW-eq}
\eea
which follows from $G^\alpha_\beta$ component of eq. (\ref{GFE})
using eqs. (\ref{metric-decomp},\ref{Tab},\ref{fluid-sum}).
The generalized nature of the gravity appears in the $F$ terms: 
one in the damping term and the other in modulating the amplitude of 
the fluid source term. 
This equation is valid for the general theory in eq. (\ref{Lagrangian}),
and the presence of arbitrary number of the minimally coupled scalar fields 
(with general $g_{IJ}$) does not formally affect the equation 
for the cosmological gravitational wave.
The presence of kinetic components
additionally introduces the corresponding Boltzmann equations,
and contributes to the anisotropic pressure in the above equations,
see \S \ref{sec:Kinetic} and \ref{sec:CMB}.

Equation (\ref{GW-eq}) can be arranged in following form
\bea
   & & v_t^{\prime\prime} + \left( k^2 + 2 K - {z_t^{\prime\prime} \over z_t}
       \right) v_t = {a^3 \over \sqrt{F}} \sum_{l} \pi^{(t)}_{(l)},
   \nonumber \\
   & & \quad
       v_t \equiv a \sqrt{F} c^{(t)}, \quad
       z_t \equiv a \sqrt{F},
   \label{GW-v-eq}
\eea
where a prime denotes the time derivative based on $\eta$.
In the large-scale limit, thus ignoring the $k^2$ term in eq. (\ref{GW-v-eq}), 
and assuming $K = 0$ and $\pi^{(t)} = 0$,
we have a general integral form solution \cite{Hwang-1991-PRW}
\bea
   & & c^{(t)} (k, t) = c (k) - d (k) \int^t {1 \over a^3 F} dt,
   \label{GW-sol2}
\eea 
where $c(k)$ and $d(k)$ are integration constants
for relatively growing and decaying solutions, respectively.
This solution is valid considering the general time evolution of 
the background dynamics as long as the perturbation is in the superhorizon. 
The growing solution is simply {\it conserved} in the superhorizon scale and 
the generalized nature of the gravity does not affect the conserved 
nature of the growing solution.
Only in the decaying solution the generalized gravity nature appears 
explicitly.

Similar equations and solutions as above can be derived for a single 
component scalar-type perturbation in unified forms for the fluid,
the field, and the generalized gravity theory as well
\cite{Noh-Hwang-2001-Unified}.

\section{Kinetic Theory Formulation}
                                       \label{sec:Kinetic}

\subsection{Relativistic Boltzmann equation}
                                       \label{sec:Boltzmann-eq}

The evolutions of collisionless particles and the photon are
described by specifying distribution functions
which are governed by the corresponding Boltzmann equations.
The relativistic Boltzmann equation is given as 
\cite{Lindquist-1966-etal}
\bea
   {d \over d \lambda} f 
   &=& {d x^a \over d \lambda}
       {\partial f \over \partial x^a} + {d p^a \over d \lambda}
       {\partial f \over \partial p^a}
       = p^a {\partial f \over \partial x^a}
       - \Gamma^a_{bc} p^b p^c {\partial f \over \partial p^a}
   \nonumber \\
   &=& C [f],
   \label{Boltzmann-eq}
\eea
where $f (x^a, p^b)$ is a distribution function with the phase space
variables $x^a$ and $p^a \equiv d x^a/d\lambda$,
and $C[f]$ is the collision term.
The energy-momentum tensor of the kinetic component with mass $m$ is given as
\bea
   & & T_{(c)}^{ab} = \int 2 \theta (p^0) \delta (p^c p_c + m^2 ) 
       p^a p^b f \sqrt{-g} d^4 p^{0123}.
   \label{Tab-Kin-1}
\eea
Assuming the mass-shell condition, after integrating over $p^0$, we have
\bea
   & & T_{(c)}^{ab} = \int {\sqrt{-g} d^3 p^{123} \over |p_0|} p^a p^b f.
   \label{Tab-Kin}
\eea
Equations (\ref{GFE},\ref{EOM},\ref{fluid-conservation}) 
together with eqs. (\ref{Boltzmann-eq},\ref{Tab-Kin}),
including $T_{(c)ab}$ in the individual fluid energy-momentum tensor,
provide a complete set of equations considering the 
contribution of a component based on the distribution function 
(we call it a kinetic component).
The corresponding fluid quantities can be identified using eq. (\ref{Tab-f}).
In the case of multiple kinetic components, we have
eqs. (\ref{Boltzmann-eq},\ref{Tab-Kin}) now valid for the individual
kinetic component.
The corresponding fluid quantities of the individual component can be 
identified using eqs. (\ref{Tab-f},\ref{Tab-sum}).

\subsection{Boltzmann equation in perturbed FLRW}
                                       \label{sec:perturbed-Boltzmann}

Under the perturbed FLRW metric in eq. (\ref{metric-general}), 
using $p^a$ as the phase space variable, eq. (\ref{Boltzmann-eq}) becomes
\bea
   & & p^0 f^\prime + p^\alpha f_{,\alpha} - \Bigg[ {a^\prime \over a} \left(
       p^0 p^0 + g^{(3)}_{\alpha\beta} p^\alpha p^\beta \right)
       + A^\prime p^0 p^0 
   \nonumber \\
   & & \quad
       + 2 \left( A_{,\alpha} - {a^\prime \over a} B_\alpha
       \right) p^0 p^\alpha
   \nonumber \\
   & & \quad
       + \left( - 2 {a^\prime \over a} g^{(3)}_{\alpha\beta} A
       + B_{\alpha|\beta}
       + C^\prime_{\alpha\beta} + 2 {a^\prime \over a} C_{\alpha\beta} \right)
       p^\alpha p^\beta \Bigg] {\partial f \over \partial p^0}
   \nonumber \\
   & & \quad
       - \left( 2 {a^\prime \over a} p^0 p^\alpha
       + \Gamma^{(3)\alpha}_{\beta\gamma} p^\beta p^\gamma \right)
       {\partial f \over \partial p^\alpha} = C[f].
   \label{f-eq}
\eea
In handling the Boltzmann equation and the energy-momentum tensor 
in perturbed FLRW spacetime, it is convenient to introduce a special
phase space variables based on a tetrad frame.
In literature we find several different choices for the phase space variables
\cite{tetrad,Peebles-1982,Bond-Szalay-1983,Kodama-Sasaki-1984}.
As the phase space variables we use $(q, \gamma^\alpha)$ introduced as
\bea
   & & p^0 \equiv {1 \over a^2} ( 1 - A ) \sqrt{q^2 + m^2 a^2}, \quad
   \nonumber \\
   & & p^\alpha \equiv {1 \over a^2} \left( q \gamma^{\alpha}
       + \sqrt{q^2 + m^2 a^2} B^\alpha
       - q \gamma^{\beta} C^\alpha_\beta \right),
   \label{q-def}
\eea
where $\gamma^\alpha$ is based on $g^{(3)}_{\alpha\beta}$
with $\gamma^\alpha \gamma_\alpha = 1$.
The advantage of this choice in our gauge-ready approach will become
clear below eq. (\ref{GT-f}).
Using $(q, \gamma^\alpha)$ as the phase space variables 
eq. (\ref{f-eq}) becomes
\bea
   & & f^\prime 
       + {q \over \sqrt{q^2 + m^2 a^2} } \left( \gamma^\alpha f_{,\alpha}
       - \Gamma^{(3)\alpha}_{\;\;\;\;\;\beta\gamma}
       \gamma^\beta \gamma^\gamma {\partial f \over \partial \gamma^\alpha}
       \right)
   \nonumber \\
   & & \quad
       - \left[ {\sqrt{q^2 + m^2 a^2} \over q}
       A_{,\alpha} \gamma^\alpha + \left( B_{\alpha|\beta}
       + C_{\alpha\beta}^\prime \right) \gamma^\alpha \gamma^\beta
       \right] q {\partial f \over \partial q}
   \nonumber \\
   & & \quad 
       = {a^2 \over \sqrt{q^2 + m^2 a^2} } (1 + A) C [f].
   \label{Boltzmann-eq-q}
\eea
We decompose the distribution function into the background and 
the perturbed order as
\bea
   & & f (\eta, x^\alpha, q, \gamma^\alpha )
       = \bar f (\eta, q) + \delta f (\eta, x^\alpha, q, \gamma^\alpha ).
   \label{f-decomp}
\eea
Assuming that the collision term has no role to the background order, 
in which the Thomson scattering is the case, we have
\bea
   & & \bar f^\prime = 0.
   \label{BG-f-eq}
\eea
Thus, $\bar f$ is a function of $q$ only.
The energy-momentum tensor in eq. (\ref{Tab-Kin}) becomes
\bea
   & & T_{(c)}^{ab}
       = {1 \over a^2} \int p^a p^b f {q^2 d q d \Omega_q
       \over \sqrt{ q^2 + m^2 a^2 } }.
\eea
The fluid quantities defined in eq. (\ref{Tab-decomp}) become:
\bea
   & & \mu_{(c)} = {1 \over a^4} \int f \sqrt{q^2 + m^2 a^2} q^2 dq d \Omega_q,
   \nonumber \\
   & & p_{(c)} = {1\over 3 a^4} \int f {q^4 dq d \Omega_q
       \over \sqrt{q^2 + m^2 a^2} },
   \nonumber \\
   & & ( \mu_{(c)} + p_{(c)} ) v_{(c)\alpha}
       = {1 \over a^4} \int \delta f \gamma_\alpha q^3 dq d \Omega_q,
   \nonumber \\
   & & \pi_{(c)\beta}^{(3)\alpha}
       = {1 \over a^4} \int \delta f \left( \gamma^\alpha \gamma_\beta
       - {1\over 3} \delta^\alpha_\beta \right) {q^4 dq d \Omega_q \over
       \sqrt{q^2 + m^2 a^2} }.
   \label{Fluid-f}
\eea

Under the gauge transformation $\tilde x^a = x^a + \xi^a$,
considering $p^a \equiv d x^a/ d \lambda$, we have
$\tilde p^a = p^a + \xi^a_{\;\;,b} p^b$.
Using the definition of $q$ in eq. (\ref{q-def}), and
using eq. (\ref{GT-1}) we have
\bea
   & & \tilde q = q + q H \xi^t + \sqrt{q^2 + m^2 a^2} {1 \over a}
       \xi^t_{,\alpha} \gamma^\alpha.
   \label{GT-q}
\eea
Since $\bar f$ depends only on $q$, we have
$\delta \tilde f = \delta f - (\partial \bar f/ \partial q) (\tilde q - q)$,
thus
\bea
   & & \delta \tilde f = \delta f - q {\partial f \over \partial q}
       \left( H \xi^t + {\sqrt{q^2 + m^2 a^2} \over q} {1 \over a}
       \xi^t_{\;,\alpha} \gamma^\alpha \right).
   \label{GT-f}       
\eea
Notice that with our phase space variables in eq. (\ref{q-def})
the perturbed distribution function $\delta f$ is spatially gauge-invariant.
Thus, our choice of the phase space variables is particularly convenient 
for the gauge-ready formulation where, as a strategy for later convenient 
use, we do not fix the temporal gauge condition while fixing the
spatial gauge condition without losing any advantage; our
$\delta f$ is spatially gauge invariant.
We can show that the gauge transformation property of $\delta f$
is consistent with the gauge transformation properties
of the fluid quantities identified in eq. (\ref{Fluid-f}).

\subsection{Background equations}
                                          \label{sec:BG-evol}

Equations (\ref{BG1}-\ref{BG6}) describe the evolution of 
the FLRW world model.
The sum over fluid quantities in eq. (\ref{fluid-sum}) should include
the kinetic components.  
To the background order, from eq. (\ref{Fluid-f}) we have
\bea
   & & \mu_{(c)} = {4 \pi \over a^4} \int f \epsilon q^2 dq, \quad
       p_{(c)} = {4 \pi \over 3 a^4} \int f {q^4 \over \epsilon} dq,
   \label{fluid-massive-BG}
\eea
where $\epsilon (q,\eta) \equiv \sqrt{q^2 + m^2 a^2}$;
hereafter, the mass $m$ appears only in $\epsilon$.
Thus, for massless particles we have
\bea
   & & \mu_{(c)} = {4 \pi \over a^4} \int f q^3 dq, \quad
       p_{(c)} = {1 \over 3} \mu_{(c)}.
   \label{fluid-massless-BG}
\eea
We can show that eq. (\ref{BG5}) applies to the kinetic components as well
with $(i) = (c)$ both for the massless and massive particles.
This identification gives
\bea
   & & Q_{(c)} = 0.
   \label{Q-BG}
\eea

In the case we have the matter ($m$), radiation ($r$), 
and massive neutrino ($\nu_m$), it is convenient to introduce
\bea
   & & \Omega_m \equiv {\mu_m \over \mu_c}, \quad
       \Omega_r \equiv {\mu_r \over \mu_c}, \quad
       \Omega_{\nu_m} \equiv {\mu_{\nu_m} \over \mu_c}, \quad
       \Omega_{K} \equiv - {K \over a^2 H^2},
   \nonumber \\
   & & \Omega_{\Lambda} \equiv {\Lambda \over 3 H^2}, \quad
       \mu_c \equiv {3 H^2 \over 8 \pi G}.
   \label{Omega-defs}
\eea
The matter includes the baryon and the CDM 
with $\Omega_m = \Omega_b + \Omega_C$,
and the radiation includes photons and the massless collisionless particles
like massless neutrino ($\nu$) with $\Omega_r = \Omega_\gamma + \Omega_\nu$.
In such a case we have
\bea
   & & \mu = \sum_{l} \mu_{(l)} = \mu_b + \mu_C + \mu_\gamma
       + \mu_\nu + \mu_{\nu_m} + \mu_\Lambda, 
   \nonumber \\
   & & p = \sum_{l} p_{(l)} = p_\gamma + p_\nu + p_{\nu_m} + p_\Lambda,
   \label{mu-p-sum}
\eea
where we have recovered the cosmological constant using the
prescription below eq. (\ref{BG6}).

The distribution function of the fermi/bose ($\pm$ sign) particle is given as
\bea
   & & f (\epsilon) = {g_s \over h_P^3} {1 \over e^{\epsilon/(k_B a T)} \pm 1},
   \label{f-BG}
\eea 
where $g_s$ is the number of spin degrees of freedom, and $h_P$ and $k_B$ 
are the Planck and the Boltzmann constants.
If the decoupling of the massive particle occurs while it is relativistic,
thus for neutrino mass much less than 1MeV, $\epsilon$ in eq. (\ref{f-BG})
can be approximated as $q$, and afterward the distribution function 
is well approximated by Fermi-Dirac distribution with zero rest mass.

Since $Q_{(i)} = 0$, from eq. (\ref{BG5})
we have $\mu_m \equiv \mu_b + \mu_C \propto a^{-3}$ and
$\mu_r \equiv \mu_\gamma + \mu_\nu \propto a^{-4}$.
{}For the photon and massless neutrinos we have
\bea
   & & p_\gamma = {1 \over 3} \mu_\gamma, \quad
       p_\nu = {1 \over 3} \mu_\nu, \quad
       \mu_\nu = N_\nu {7 \over 8} \left( {4 \over 11} \right)^{4/3} 
       \mu_\gamma, 
   \\
   & & T_\nu = \left( {4 \over 11} \right)^{1/3} T_\gamma,
\eea
where $N_\nu$ is the number of massless neutrino species.

Equations (\ref{BG1}-\ref{BG5}) together with 
eqs. (\ref{Q-BG},\ref{mu-p-sum}) describe the background
evolution in the context of generalized gravity theories.
The generalization to include multi-component massive/massless
collisionless components is trivial: we simply consider the fluid 
quantities in eqs. (\ref{fluid-massive-BG},\ref{fluid-massless-BG}) 
for each component.

\subsection{Massless particle}
                                       \label{sec:massless}

{}For a massless particle, $m = 0$, it is convenient to
introduce a frequency integrated perturbed intensity
\bea
   & & \delta I (x^a, \hat {\bf \gamma}) 
       \equiv 4 \Theta (x^a, \hat {\bf \gamma})
       \equiv {\int \delta f q^3 dq \over \int \bar f q^3 dq}.
\eea
{}For photons, unless we have an energy injection process into the
CMB, the spectral distortion vanishes to the linear order.
Assuming photon's distribution function
\bea
   & & f = {g_s \over h_P^3} {1 \over e^{q/ (k_B a_0 T_0)} -1},
\eea
we can expand the temperature fluctuation $\Theta \equiv \delta T/ T$ 
in the following form as well
\bea
   & & \Theta ( x^a, q , \hat \gamma^\alpha ) 
       \equiv {\delta f \over - q {\partial f \over \partial q} }.
\eea
In terms of $\Theta$ the perturbed part of eq. (\ref{Boltzmann-eq-q}) becomes
\bea
   & & \Theta^\prime
       + \left( \gamma^\alpha {\partial \over \partial x^\alpha}
       - \Gamma^{(3)\alpha}_{\;\;\;\;\;\beta\gamma} \gamma^\beta \gamma^\gamma
       {\partial \over \partial \gamma^\alpha} \right) \Theta
   \nonumber \\
   & & \quad
       = - \gamma^\alpha A_{,\alpha}
       - \left( B_{\alpha|\beta} + C_{\alpha\beta}^\prime \right)
       \gamma^\alpha \gamma^\beta
   \nonumber \\
   & & \quad
       + {\rm collision \; term.}
   \label{Theta-eq}
\eea
The fluid quantities in eq. (\ref{Fluid-f}) become
\bea
   & & {\delta \mu_{(c)} \over \bar \mu_{(c)}}
       = 4 \int \Theta {d \Omega_q \over 4 \pi}, \quad
       \delta p_{(c)} = {1 \over 3} \delta \mu_{(c)},
   \nonumber \\
   & & v_{(c)\alpha}
       = 3 \int \Theta \gamma_\alpha {d \Omega_q \over 4 \pi},
   \nonumber \\
   & & \pi^{(3)\alpha}_{(c)\beta} = 4 \mu_{(c)} \int \Theta
       \left( \gamma^\alpha \gamma_\beta - {1 \over 3} \delta^\alpha_\beta
       \right) {d \Omega_q \over 4 \pi}.
   \label{fluid-massless}
\eea

Using the spatial and momentum harmonic function introduced by Hu {\it et al} 
in \cite{Hu-etal-1998}, we expand
\bea
   & & \Theta ({\bf x}, \eta, \hat {\bf \gamma})
       \equiv \sum_{\bf k} \sum_{l = 0}^\infty \sum_{m = -2}^2
       \Theta_{(\ell)}^{(m)} ({\bf k}, \eta)
       G_{(\ell)}^{(m)} ({\bf k}; {\bf x}, \hat {\bf \gamma}),
   \nonumber \\
   \label{Theta-decomp} \\
   & & G_{(\ell)}^{(m)} ({\bf k}; {\bf x}, \hat {\bf \gamma}) \equiv
       (-i)^\ell \sqrt{ 4 \pi \over 2 \ell + 1}
       Y^m_\ell ({\bf k}, \hat {\bf \gamma})
       e^{i \delta ({\bf k}; {\bf x})}, 
   \label{G-def}
\eea
where $\ell \ge |m|$, and $m = 0, \; \pm 1, \; \pm 2$ correspond to the
scalar-, vector-, and tensor-type perturbations, respectively.
Thus, for the scalar-type perturbation we have
\bea
   & & G^{(0)}_{(\ell)} = (-i)^\ell
       P_\ell (\hat {\bf k} \cdot \hat \gamma)
       e^{i \delta ({\bf k}; \cdot {\bf x})},
\eea
where $\delta ({\bf k}; {\bf x})$ is a spatially dependent phase factor
which depends on the harmonic functions, see eq. (\ref{G-Y-def});
in the flat background we have 
$e^{i \delta ({\bf k}; {\bf x})} = e^{i {\bf k} \cdot {\bf x}}$.
As the normalization we have
\bea
   & & \int \left| G_{(\ell)}^{(m)} G_{(\ell^\prime)}^{(m^\prime) *} \right|
       {d \Omega_p \over 4 \pi} = {1 \over 2 \ell + 1}
       \delta_{\ell \ell^\prime} \delta_{m m^\prime}.
\eea
We have the recursion relation \cite{Hu-etal-1998}
\bea
   & & G_{(\ell) | \alpha}^{(m)} \gamma^\alpha 
       \equiv \left( \gamma^\alpha {\partial \over \partial x^\alpha}
       - \Gamma^{(3)\alpha}_{\;\;\;\;\;\beta\gamma}
       \gamma^\beta \gamma^\gamma {\partial \over \partial \gamma^\alpha}
       \right) G_{(\ell)}^{(m)}
   \nonumber \\
   & & \quad
       = {k \over 2 \ell + 1} \left( \kappa_\ell^m G_{(\ell -1)}^{(m)}
       - \kappa_{\ell + 1}^m G_{(\ell + 1)}^{(m)} \right),
   \label{G-recursion}
\eea
where $\kappa_0^0 \equiv 1$ and for $\ell \ge 1$
\bea
   & & \kappa_\ell^m \equiv \sqrt{ (\ell^2 - m^2)
       \left[ 1 - ( \ell^2 - 1 - |m| ) {K \over k^2} \right] }.
   \label{kappa-def}
\eea

The harmonic functions are introduced such that we have
$(0)$, $(\pm 1)$, and $(\pm 2)$ superscripts instead of
$(s)$, $(v)$, and $(t)$ indices introduced in eq. (\ref{Y-def}).
In terms of the spatial harmonic functions we identify
\bea
   & & G^{(m)}_{(|m|)} \equiv \gamma^{\alpha_1} \dots \gamma^{\alpha_{|m|}}
       Y^{(m)}_{\alpha_1 \dots \alpha_{|m|}}.
   \label{G-Y-def}
\eea
Then, from eqs. (\ref{G-recursion},\ref{Y-def}) we can show:
\bea
   & & G_{(0)}^{(0)} \equiv Y^{(0)}, \quad
       G_{(1)}^{(0)} = \gamma^{\alpha} Y^{(0)}_{\alpha}, 
   \nonumber \\
   & & {2 \over 3} \sqrt{ 1 - 3 {K \over k^2} } G_{(2)}^{(0)}
       = \gamma^{\alpha} \gamma^{\beta} Y^{(0)}_{\alpha\beta};
   \nonumber \\
   & & G_{(1)}^{(\pm 1)} \equiv \gamma^{\alpha} Y^{(\pm 1)}_{\alpha}, \quad
       {1 \over \sqrt{3}} \sqrt{ 1 - 2 {K \over k^2} } G_{(2)}^{(\pm 1)}
       = \gamma^{\alpha} \gamma^{\beta} Y^{(\pm 1)}_{\alpha \beta};
   \nonumber \\
   & & G_{(2)}^{(\pm 2)} \equiv \gamma^{\alpha} \gamma^{\beta}
       Y^{(\pm 2)}_{\alpha \beta}.
   \label{G-Y}
\eea

Now, from eqs. (\ref{Theta-eq}) using the expansion in 
eqs. (\ref{Theta-decomp},\ref{G-def}) and the recursion relation
in eq. (\ref{G-recursion}) we can derive
\bea
   & & \dot \Theta_{(\ell)}^{(m)}
       = {k \over a} \left( {1 \over 2 \ell -1} \kappa_\ell^m 
       \Theta_{(\ell -1)}^{(m)} - {1 \over 2 \ell + 3} \kappa^m_{\ell + 1}
       \Theta_{(\ell +1)}^{(m)} \right)
   \nonumber \\
   & & \quad
       + M_{(\ell)}^{(m)} + C_{(\ell)}^{(m)},
   \label{Theta-lm-eq}
\eea
where $M_{(\ell)}^{(m)}$ and $C_{(\ell)}^{(m)}$ are the metric perturbation
and the collision term, respectively.
The collision terms for the Thomson scattering in photon distribution
function together with polarizations will be considered in 
\S \ref{sec:scattering}.
The metric perturbations in eq. (\ref{Theta-eq}) can be calculated using
eqs. (\ref{metric-decomp},\ref{G-Y}) as
\bea
   & & M_{(\ell)}^{(m)}
       = \left(
       \begin{array}{ccc}
       - \dot \varphi + {k^2 \over 3 a^2} \chi & {k \over a} \alpha &
       - {2 \over 3} \sqrt{ 1 - 3 {K \over k^2} } {k^2 \over a^2} \chi \\
       0 & 0 &  
       {1 \over \sqrt{3}} \sqrt{ 1 - 2 {K \over k^2} } 
       {k \over a} \Psi^{(\pm 1)} \\
       0 & 0 & - \dot c^{(\pm 2)}
       \end{array}
       \right),
   \nonumber \\
   \label{M-matrix}
\eea
where the rows indicate $m = 0, \pm 1, \pm 2$, and the columns indicate
$l = 0, 1, 2$, respectively.
Using eq. (\ref{Theta-decomp}) the perturbed order fluid quantities 
in eq. (\ref{fluid-massless}) become:
\bea
   & & \delta_{(c)} \equiv {\delta \mu_{(c)} \over \mu_{(c)}}
       = 4 \Theta_{(0)}^{(0)}, \quad
       \delta p_{(c)} = {1 \over 3} \delta \mu_{(c)}, 
   \nonumber \\
   & & v_{(c)} = \Theta_{(1)}^{(0)}, \quad
       {\pi_{(c)}^{(0)} \over \mu_{(c)}}
       = {4 \over 5} {1 \over \sqrt{1 - 3 {K \over k^2} }} \Theta_{(2)}^{(0)}, 
   \nonumber \\
   & & v^{(\pm 1)}_{(c)} = \Theta_{(1)}^{(\pm 1)}, \quad
       {\pi_{(c)}^{(\pm 1)} \over \mu_{(c)}}
       = {8 \over 15} {\sqrt{3} \over \sqrt{1 - 2 {K \over k^2} }}
       \Theta_{(2)}^{(\pm 1)},
   \nonumber \\
   & & {\pi_{(c)}^{(\pm 2)} \over \mu_{(c)}}
       = {8 \over 15} \Theta_{(2)}^{(\pm 2)},
   \label{fluid-Theta}
\eea
where we used eqs. (\ref{fluid-decomp},\ref{Tab},\ref{G-Y}).

Under the gauge transformation, from eq. (\ref{GT-f}) we have
\bea
   & & \tilde \Theta = \Theta + H \xi^t 
       + {1 \over a} \xi^t_{\;,\alpha} \gamma^\alpha.
\eea
By expanding $\xi^t = \sum_{\bf k} \xi^t Y^{(0)}$ we have
\bea
   & & \tilde \Theta_{(0)}^{(0)} = \Theta_{(0)}^{(0)} + H \xi^t, \quad
       \tilde \Theta_{(1)}^{(0)} = \Theta_{(1)}^{(0)} - {k \over a} \xi^t, 
   \label{GT-Theta}
\eea
and other $\Theta^{(m)}_{(\ell)}$ are gauge-invariant.
These are consistent with the identifications in 
eq. (\ref{fluid-Theta}) and the gauge transformation properties
in eqs. (\ref{GT-1},\ref{GT-i-1}).
Thus, the temporal gauge conditions fixing $\Theta_{(0)}^{(0)}$
and $\Theta_{(1)}^{(0)}$ can be considered as belonging to
eq. (\ref{Gauges-2}).

\subsection{Massive collisionless particle}
                                       \label{sec:massive}

{}For massive collisionless particles the perturbed Boltzmann equation in 
eq. (\ref{Boltzmann-eq-q}) becomes
\bea
   & & \delta f^\prime
       + {q \over \epsilon} \left( \gamma^\alpha \delta f_{,\alpha}
       - \Gamma^{(3)\alpha}_{\;\;\;\;\;\beta\gamma} \gamma^\beta \gamma^\gamma
       {\partial \delta f \over \partial \gamma^\alpha} \right)
       - \Bigg[ {\epsilon \over q}
       A_{,\alpha} \gamma^\alpha 
   \nonumber \\
   & & \quad
       + \left( B_{\alpha|\beta}
       + C_{\alpha\beta}^\prime \right) \gamma^\alpha \gamma^\beta
       \Bigg] q {\partial \bar f \over \partial q} = 0.
   \label{Boltzmann-eq-pert-m}
\eea
In the massive case we can expand directly the perturbed distribution function.
Instead of $\delta f$ we use the following variable
\bea
   & & \hat \Theta ({\bf x}, \eta, q, \hat {\bf \gamma})
       \equiv {\delta f \over - q {\partial f \over \partial q}}
   \nonumber \\
   & & \quad
       \equiv \sum_{\bf k} \sum_{l = 0}^\infty \sum_{m = -2}^2
       \hat \Theta_{(\ell)}^{(m)} ({\bf k}, \eta, q)
       G_{(\ell)}^{(m)} ({\bf k}; {\bf x}, \hat {\bf \gamma}).
   \label{hat-Theta-decomp}
\eea
{}From eq. (\ref{Boltzmann-eq-pert-m}), using 
eqs. (\ref{hat-Theta-decomp},\ref{G-recursion},\ref{metric-decomp},\ref{G-Y}) 
we can derive
\bea
   & & \dot {\hat \Theta}_{(\ell)}^{(m)}
       = {q \over \epsilon} {k \over a} 
       \left( {1 \over 2 \ell -1} \kappa_\ell^m \hat \Theta_{(\ell -1)}^{(m)}
       - {1 \over 2 \ell + 3} \kappa^m_{\ell + 1}
       \hat \Theta_{(\ell +1)}^{(m)} \right)
   \nonumber \\
   & & \quad
       + \hat M_{(\ell)}^{(m)},
   \label{hat-Theta-lm-eq} \\
   & & \hat M_{(\ell)}^{(m)}
       = \left(
       \begin{array}{ccc}
       - \dot \varphi + {k^2 \over 3 a^2} \chi 
       & {\epsilon \over q} {k \over a} \alpha &
       - {2 \over 3} \sqrt{ 1 - 3 {K \over k^2} } {k^2 \over a^2} \chi \\
       0 & 0 &  {1 \over \sqrt{3}} \sqrt{ 1 - 2 {K \over k^2} } 
       {k \over a} \Psi^{(\pm 1)} \\
       0 & 0 & - \dot c^{(\pm 2)}
       \end{array}
       \right),
   \nonumber \\
   \label{hat-M-matrix}
\eea
where $\ell \ge |m|$.
The rows and columns of $\hat M_{(\ell)}^{(m)}$ indicate $m = 0, \pm 1, \pm 2$,
and $l = 0, 1, 2$, respectively.
In the massless limit $\hat M_{(\ell)}^{(m)}$ reduces to 
$M_{(\ell)}^{(m)}$ in eq. (\ref{M-matrix}), and eq. (\ref{hat-Theta-lm-eq})
reduces to eq. (\ref{Theta-lm-eq}).
Thus, we can regard eqs. (\ref{hat-Theta-lm-eq},\ref{hat-M-matrix})
as being valid for both massless and massive collisionless particles.

{}From eqs. (\ref{Fluid-f},\ref{hat-Theta-decomp}) 
the perturbed order fluid quantities become:
\bea
   & & \delta \mu_{(c)} = {4 \pi \over a^4} \int 
       \hat \Theta^{(0)}_{(0)} \left( - {\partial f \over \partial q} \right)
       \epsilon q^3 dq, 
   \nonumber \\
   & & \delta p_{(c)} = {4 \pi \over 3 a^4} \int 
       \hat \Theta^{(0)}_{(0)} \left( - {\partial f \over \partial q} \right)
       {q^5 \over \epsilon} dq,
   \nonumber \\
   & & ( \mu_{(c)} + p_{(c)} ) v^{(m)}_{(c)}
       = {4 \pi \over 3 a^4} \int 
       \hat \Theta_{(1)}^{(m)} \left( - {\partial f \over \partial q} \right)
       q^4 dq, 
   \nonumber \\
   & & \hskip 5cm
       (m = 0, \pm 1), 
   \nonumber \\
   & & \pi^{(0)}_{(c)} = {1 \over \sqrt{1 - 3 {K \over k^2} } }
       {4 \pi \over 5 a^4} \int 
       \hat \Theta^{(0)}_{(2)} \left( - {\partial f \over \partial q} \right) 
       {q^5 \over \epsilon} dq, 
   \nonumber \\
   & & \pi^{(\pm 1)}_{(c)}
       = {\sqrt{3} \over \sqrt{1 - 2 {K \over k^2} } }
       {8 \pi \over 15 a^4} \int 
       \hat \Theta^{(\pm 1)}_{(2)} 
       \left( - {\partial f \over \partial q} \right)
       {q^5 \over \epsilon} dq, 
   \nonumber \\
   & & \pi^{(\pm 2)}_{(c)}
       = {8 \pi \over 15 a^4} \int \hat \Theta^{(\pm 2)}_{(2)} 
       \left( - {\partial f \over \partial q} \right) {q^5 \over \epsilon} dq,
   \label{fluid-massive}
\eea
where we used eqs. (\ref{fluid-decomp},\ref{Tab},\ref{G-Y}).
In the massless limit, assuming $\hat \Theta^{(m)}_{(\ell)}$ is
independent of $q$,
the fluid quantities in eqs. (\ref{fluid-massive-BG},\ref{fluid-massive}) 
reduce to the ones in the massless case 
eqs. (\ref{fluid-massless-BG},\ref{fluid-Theta}) with
$\hat \Theta^{(m)}_{(\ell)} = \Theta^{(m)}_{(\ell)}$.

Under the gauge transformation, using eq. (\ref{GT-f}) we have
\bea
   & & \tilde {\hat \Theta} = \hat \Theta + H \xi^t 
       + {\epsilon \over q} {1 \over a} \xi^t_{\;,\alpha} \gamma^\alpha.
\eea
Thus,
\bea
   & & \tilde {\hat \Theta}_{(0)}^{(0)} 
       = \hat \Theta_{(0)}^{(0)} + H \xi^t, \quad
       \tilde {\hat \Theta}_{(1)}^{(0)} = \hat \Theta_{(1)}^{(0)} 
       - {\epsilon \over q} {k \over a} \xi^t, 
   \label{GT-hat-Theta}
\eea
and other $\hat \Theta^{(m)}_{(\ell)}$ are gauge-invariant.
These are consistent with the identifications in
eq. (\ref{fluid-massive}) and the gauge transformation properties
in eqs. (\ref{GT-1},\ref{GT-i-1}).
Thus, the temporal gauge conditions fixing $\hat \Theta_{(0)}^{(0)}$
and $\hat \Theta_{(1)}^{(0)}$ can be considered as belonging to
eq. (\ref{Gauges-2}).

Thus, we have complete sets of perturbation equations for three-types
of perturbations including a single component massive collisionless particle.
The eq. (\ref{hat-Theta-lm-eq}) together with the gravitational field 
equations in eqs. (\ref{G1}-\ref{G10},\ref{rot-1}-\ref{rot-i},\ref{GW-eq}),
and the fluid quantities for the massive particle in 
eq. (\ref{fluid-massive}) provide the complete sets.
In the case of multi-component massive collisionless particles 
we simply consider eqs. (\ref{hat-Theta-lm-eq},\ref{fluid-massive})
for each component of massive collisionless particle.
The equations for the scalar-type perturbation are designed in 
a gauge-ready form.
The collective (or the sum over individual) fluid quantities in 
eqs. (\ref{G2}-\ref{G5},\ref{G7},\ref{rot-2},\ref{GW-eq})
include the kinetic components, whereas $(i)$ in 
eqs. (\ref{G9},\ref{G10},\ref{rot-i}) does not include the kinetic components.
Using eqs. (\ref{fluid-massive},\ref{Q-BG}), however,
we can show that eq. (\ref{hat-Theta-lm-eq})
gives eqs. (\ref{G9},\ref{G10},\ref{rot-i}) with $(i) = (c)$.
This identification gives 
\bea
   & & \delta Q_{(c)} = 0 = J_{(c)}^{(s,v)}.
\eea
This follows because we have assumed collisionless situation
(and with no direct interaction between the kinetic component
and the other components).
{}For the case of photon with Thomson scattering, see
eq. (\ref{interaction-Thomson}).

\section{CMB Anisotropy}
                                       \label{sec:CMB}

\subsection{Thomson scattering and polarizations}
                                       \label{sec:scattering}

Besides the photon distribution function for the temperature 
(or total intensity) fluctuation $f = f_\Theta$, we have three other photon 
distribution functions describing the state of polarization,
$f_Q$, $f_U$, and $f_V$.
$\Theta$, $Q$, $U$, and $V$ form the four Stokes parameters. 
We will ignore the fourth Stokes parameter $V$ describing a circular 
polarization because it cannot be generated through Thomson scattering 
in standard FLRW cosmological models.
While the temperature behaves as a scalar quantity $Q$ and $U$ do not.
It is known that the combinations $Q \pm i U$ behave as the spin $\pm 2$
quantities \cite{Zaldarriaga-Seljak-1997}. 
Thus, while $\Theta$ can be expanded in ordinary spherical harmonic
$Y_{lm}$, $Q \pm i U$ should be expanded in the spin-weighted harmonics
${}_{\pm 2} Y_{lm}$ \cite{Newman-Penrose-1966,Dautcourt-Rose-1978}.

{}Following the convention in \cite{Hu-etal-1998} 
we expand $\Theta$ and $Q \pm i U$ in terms of the spin-weighted 
spatial and momentum harmonic functions:
\bea
   & & \Theta ({\bf x}, \eta, \hat {\bf \gamma}) 
       \equiv \sum_{\bf k} \sum_{m = -2}^2 \sum_\ell
       \Theta_{(\ell)}^{(m)} ({\bf k}, \eta) 
       {}_0 G_{(\ell)}^{(m)} ({\bf k}; {\bf x}, \hat {\bf \gamma}),
   \nonumber \\
   & & Q ({\bf x}, \eta, \hat {\bf \gamma}) 
       \pm i U ({\bf x}, \eta, \hat {\bf \gamma}) 
   \nonumber \\
   & & \quad
       \equiv \sum_{\bf k} \sum_{m = -2}^2 \sum_\ell
       \left[ E_{(\ell)}^{(m)} ({\bf k}, \eta) 
       \pm i B_{(\ell)}^{(m)} ({\bf k}, \eta) \right] 
   \nonumber \\
   & & \quad \quad \times
       {}_{\pm 2} G_{(\ell)}^{(m)} ({\bf k}; {\bf x}, \hat {\bf \gamma}),
   \label{QU-expansion}
\eea
where
\bea
   & & {}_s G_{(\ell)}^{(m)} ({\bf k}; {\bf x}, \hat {\bf \gamma}) \equiv 
       (-i)^\ell \sqrt{ 4 \pi \over 2 \ell + 1}
       {}_s Y^m_\ell ({\bf k}, \hat {\bf \gamma})
       e^{i \delta ({\bf k}; {\bf x})}, 
   \nonumber \\
\eea
with ${}_0 Y^m_\ell \equiv Y^m_\ell$; 
thus ${}_0 G_{(\ell)}^{(m)} = G_{(\ell)}^{(m)}$.
We have a recursion relation \cite{Hu-etal-1998}
\bea
   & & {}_s G_{(\ell)|\alpha}^{(m)} \gamma^\alpha
       = {n \over 2 \ell + 1} \left( {}_s \kappa^m_\ell {}_s G^{(m)}_{(\ell -1)}
       - {}_s \kappa^m_{\ell + 1} {}_s G_{(\ell+1)}^{(m)} \right)
   \nonumber \\
   & & \quad
       - i {m n s \over \ell ( \ell +1)} {}_s G^{(m)}_{(\ell)},
   \label{G-recursion-general}
\eea
where 
\bea
   & & {}_s \kappa_\ell^m \equiv \sqrt{ {(\ell^2 - m^2)(\ell^2 - s^2) 
       \over \ell^2} \left( 1 - {\ell^2 \over n^2} K \right) }, 
   \nonumber \\
   & & n \equiv \sqrt{k^2 + ( 1 + |m| ) K },
   \label{kappa-s-def}
\eea
thus, compared with eq. (\ref{kappa-def}) we have
$n \times {}_0 \kappa_\ell^m = k \kappa_\ell^m$.
In the hyperbolic (negative curvature) background we have the
supercurvature ($0 \le k < \sqrt{|K|}$) and subcurvature ($k > \sqrt{|K|}$) 
scales for the scalar-type perturbation;
by considering $n \ge 0$ we exclude the supercurvature scale,
\cite{Lyth-Woszczyna-1995}.
It is convenient to have ${}_{-s} Y_\ell^m = (-1)^\ell {}_s Y_\ell^m$, and
other useful relations can be found in \cite{Hu-White-1997,Hu-etal-1998}.

In terms of a notation
\bea
   & & \vec{T} \equiv \left( 
       \begin{array}{c}
       \Theta \\
       Q + i U \\
       Q - i U
       \end{array}
       \right),
\eea
the Boltzmann equation can be written as
\bea
   & & \dot {\vec T} + {1 \over a} \vec T_{|\alpha} \gamma^\alpha 
       = \vec M [\vec T] + \vec C [\vec T], 
   \label{Boltzmann-general}
\eea
where the metric perturbations and the collision terms are expanded as:
\bea
   & & \vec M [\Theta] \equiv \sum_{\bf k} \sum_{m = -2}^2 \sum_\ell 
       M_{(\ell)}^{(m)} G_{(\ell)}^{(m)}, 
   \label{M-vec} \\
   & & \vec C [\Theta] \equiv \sum_{\bf k} \sum_{m = -2}^2 \sum_\ell
       C_{(\ell)}^{(m)} G_{(\ell)}^{(m)}. 
   \label{C-Theta}
\eea
The collision and the polarization terms are not affected by the perturbed 
metric, thus $\vec M [Q \pm i U] = 0$.

The collision term is derived using the total angular momentum method
in eqs. (25,26) of \cite{Hu-etal-1998}
\bea
   & & \vec C [\vec T]
       = - \dot \tau \vec T (\Omega)
       + \dot \tau \left(
       \begin{array}{c}
        \int \Theta^\prime { d \Omega^\prime \over 4 \pi}
       + \gamma^\alpha v_{(b)\alpha} \\
       0 \\
       0
       \end{array}
       \right)
   \nonumber \\
   & & \quad
       + {1 \over 10} \dot \tau \int \sum_{m = -2}^2
       {\bf P}^{(m)} (\Omega, \Omega^\prime) \cdot \vec T (\Omega^\prime)
       d \Omega^\prime,
   \label{Collision}
\eea
where $v_{(b)\alpha}$ is the baryon's perturbed velocity variable, and
${\bf P}^{(m)}$ is given in eq. (52) of \cite{Hu-White-1997};
$\dot \tau \equiv n_e x_e \sigma_T$ where $n_e$ is the 
electron density, $x_e$ is the ionization fraction, 
and $\sigma_T$ is the Thomson cross section.
The time evolution of $n_e x_e$ is determined by the recombination history.
Using ${\bf P}^{(m)}$ in \cite{Hu-White-1997} the collision term becomes 
\bea
   & & C_{(\ell)}^{(m)} 
       = - \dot \tau \Theta_{(\ell)}^{(m)} 
       + \dot \tau \left( 
       \begin{array}{ccc}
       \Theta_{(0)}^{(0)} & v_{(b)}^{(0)} & P^{(0)} \\
       0 & v_{(b)}^{(\pm 1)} & P^{(\pm 1)} \\
       0 & 0 & P^{(\pm 2)} 
       \end{array} 
       \right),
   \label{C-Theta-matrix} \\
   & & \vec C [Q \pm i U] = - \dot \tau \sum_{\bf k} \sum_{m = -2}^2 
       \sum_\ell \Big( E_{(\ell)}^{(m)} \pm i B_{(\ell)}^{(m)} 
   \nonumber \\
   & & \quad
       + \sqrt{6} P^{(m)} \delta_{\ell 2} \Big) {}_{\pm 2} G_{(2)}^{(m)},
   \label{C-QU}
\eea
where
\bea
   & & P^{(m)} \equiv {1 \over 10} \left( \Theta_{(2)}^{(m)}
       - \sqrt{6} E_{(2)}^{(m)} \right).
\eea
{}From eqs. (\ref{Boltzmann-general},\ref{QU-expansion}), using the 
recursion relation in eq. (\ref{G-recursion-general}) 
and the collision term in eq. (\ref{C-QU}), we can show:
\bea
   & & \dot E_{(\ell)}^{(m)}
       = {n \over a} \left( {1 \over 2 \ell -1} 
       {}_2 \kappa_\ell^m E_{(\ell -1)}^{(m)}
       - {1 \over 2 \ell + 3} {}_2 \kappa_{\ell+1}^m
       E_{(\ell +1)}^{(m)} \right)
   \nonumber \\
   & & \quad
       - {n \over a} {2 m \over \ell ( \ell +1)} B_{(\ell)}^{(m)}
       - \dot \tau \left( E_{(\ell)}^{(m)}
       + \sqrt{6} P^{(m)} \delta_{\ell 2} \right),
   \label{E-eq} \\
   & & \dot B_{(\ell)}^{(m)}
       = {n \over a} \left( {1 \over 2 \ell -1} 
       {}_2 \kappa_\ell^m B_{(\ell -1)}^{(m)}
       - {1 \over 2 \ell + 3} {}_2 \kappa_{\ell+1}^m
       B_{(\ell +1)}^{(m)} \right)
   \nonumber \\
   & & \quad
       + {n \over a} {2 m \over \ell ( \ell +1)} E_{(\ell)}^{(m)}
       - \dot \tau B_{(\ell)}^{(m)},
   \label{B-eq}
\eea
where $\ell \ge 2$ and $m \ge 0$, and we have $B_\ell^{(0)} = 0$.
Notice that with an identification 
$B_{(\ell)}^{(-|m|)} = - B_{(\ell)}^{(|m|)}$, $E_{(\ell)}^{(\pm |m|)}$ 
and $\Theta_{(\ell)}^{(\pm |m|)}$ satisfy identical equations for both signs
\cite{Hu-White-1997}.
Equations for $\Theta_{(\ell)}^{(m)}$ follow from
eqs. (\ref{Theta-lm-eq},\ref{M-matrix},\ref{C-Theta-matrix}).
Polarization properties of the CMB in perturbed FLRW world model 
have been actively studied in the literature;
some selected references are
\cite{Dautcourt-Rose-1978,%
Dautcourt-1980,%
Bond-Efstathiou-1984,%
Zaldarriaga-Seljak-1997,%
polarization-GW,%
polarization-others,%
Lewis-etal-2001}.

Now we have complete sets of equations for three-types of perturbations
including Thomson scattered photons with polarizations.
Equations (\ref{Theta-lm-eq},\ref{M-matrix},\ref{C-Theta-matrix}) 
for the intensity (temperature) and eqs. (\ref{E-eq},\ref{B-eq}) 
for the photon polarization together with the gravitational field equations
in eqs. (\ref{G1}-\ref{G10},\ref{rot-1}-\ref{rot-i},\ref{GW-eq}),
and the fluid quantities for the massless particle in 
eq. (\ref{fluid-Theta}) provide the complete sets.
{}For a massless collisionless particle eq. (\ref{Theta-lm-eq}) remains valid
with vanishing collision terms and the polarizations.
The generalization to include multi-component massless 
collisionless particles
is trivial: we simply consider eqs. (\ref{Theta-lm-eq},\ref{fluid-Theta})
for each component of massless collisionless particle.
We can also include additional multi-component massive collisionless 
particles by considering eqs. (\ref{hat-Theta-lm-eq},\ref{fluid-massive})
for each component of massive collisionless particle.
The equations are designed in a gauge-ready form.

The collective (or sum over individual) fluid quantities in 
eqs. (\ref{G2}-\ref{G5},\ref{G7},\ref{rot-2},\ref{GW-eq}) 
include the kinetic components.
Using eqs. (\ref{fluid-Theta},\ref{Q-BG}), however, we can show that
eq. (\ref{Theta-lm-eq}) gives eqs. (\ref{G9},\ref{G10},\ref{rot-i}) 
with $(i) = (c)$; we are using the index $c$ to indicate the kinetic
component.
This identification implies
\bea
   & & \delta Q_{(c)} = 0, \quad
       J^{(0)}_{(c)} = {4 \over 3} {a \over k} \mu_{(c)} 
       \dot \tau \left( v_{(c)} - v_{(b)} \right),
   \nonumber \\
   & & J^{(\pm 1)}_{(c)} = - {4 \over 3} a \mu_{(c)} \dot \tau
       \left( v_{(c)}^{(\pm 1)} - v_{(b)}^{(\pm 1)} \right).
   \label{interaction-Thomson}
\eea
Due to Thomson scattering there exists an interaction between photons
and baryons.
{}From eq. (\ref{Q-sum}) we have
$\sum_{l} \delta Q_{(l)} = \delta Q_{(b)} + \delta Q_{(\gamma)} = 0$
and $\sum_{l} J_{(l)}^{(m)} = J_{(b)}^{(m)} + J_{(\gamma)}^{(m)} = 0$.
Thus
\bea
   & & \delta Q_{(b)} = 0, \quad
       J_{(b)}^{(0)} = - {4 \over 3} {a \over k} \mu_{(\gamma)} 
       \dot \tau \left( v_{(\gamma)} - v_{(b)} \right),
   \nonumber \\
   & & J^{(\pm 1)}_{(b)} = - J_{(\gamma)}^{(\pm 1)}
       = {4 \over 3} a \mu_{(\gamma)} \dot \tau
       \left( v_{(\gamma)}^{(\pm 1)} - v_{(b)}^{(\pm 1)} \right).
   \label{J-b}
\eea

{}For the baryon ($b$) we have $p_{(b)} = 0 = \delta p_{(b)}$, thus
$w_{(b)} = 0$. 
However, we keep the sound speed of the baryon fluid which behaves as
\cite{Ma-Bertschinger-1995}
\bea
   & & c_{(b)}^2 \equiv {\dot p_{(b)} \over \dot \mu_{(b)}}
       = \left( 1 - {1 \over 3} {d \ln{T_b} \over d \ln{a} } \right)
       {k_B T_b \over \bar \mu m_H},
   \label{c_b}
\eea
where $\bar \mu$ is the mean molecular weight.
Thus, eqs. (\ref{G9},\ref{G10},\ref{G1}) become:
\bea
   & & \dot \delta_{(b)} = - {k \over a} v_{(b)} - 3 H \alpha + \kappa,
   \label{delta_b-eq} \\
   & & \dot v_{(b)} + H v_{(b)} = {k \over a} 
       \left( \alpha + c_{(b)}^2 \delta_{(b)} 
       - {J_{(b)}^{(0)} / \mu_{(b)}} \right).
   \label{v_b-eq}
\eea
{}For the cold dark matter ($C$) we additionally have
$Q_{(C)} = 0 = \delta Q_{(C)}$ and $J_{(C)}^{(0)} = 0$,
thus eqs. (\ref{G9},\ref{G10}) become:
\bea
   & & \dot \delta_{(C)} = - {k \over a} v_{(C)} - 3 H \alpha + \kappa,
   \label{CDM-delta-eq} \\
   & & \dot v_{(C)} + H v_{(C)} = {k \over a} \alpha.
   \label{CDM-v-eq}
\eea
We emphasize that compared with previous works,
besides the equations are valid in the context of generalized gravity
theories, our sets of equations in this paper are all in gauge-ready forms.

\subsection{Tight coupling era}
                                                    \label{sec:Tight}

In the early universe, the Thomson scattering term is large enough and
the baryons and the photons are tightly coupled.
{}For large value of $\dot \tau (= t_c^{-1} = \lambda_c^{-1})$
compared with $H (= t_H^{-1} = \lambda_H^{-1})$ 
and $k/a (= 2 \pi \lambda^{-1})$,
it is difficult to handle eqs. (\ref{Theta-lm-eq},\ref{v_b-eq}) numerically;
the polarizations are negligible in that stage.
In such a case, it is convenient to arrange the equations in the following way
\cite{Ma-Bertschinger-1995}.
{}From eq. (\ref{v_b-eq}) and the $\ell =1$ component of 
eq. (\ref{Theta-lm-eq}) we have
\bea
   & & \dot v_{(b)} = - {1 \over 1 + r} H v_{(b)}
       - {r \over 1 + r} \left( \dot v_{(\gamma)} - \dot v_{(b)} \right)
       + {k \over a} \Bigg[ \alpha 
   \nonumber \\
   & & \quad
       + {1 \over 1 + r} c_{(b)}^2 \delta_{(b)}
       + {r \over 1 + r} \left( {1 \over 4} \delta_{(\gamma)}
       - {1 \over 5} \kappa_2^0 \Theta^{(0)}_{(2)} \right) \Bigg],
   \label{v_b-eq-tight}
\eea
where we introduced $r \equiv {4 \over 3} \mu_{(\gamma)}/\mu_{(b)}$.
{}From the $\ell = 2$ component of eq. (\ref{Theta-lm-eq}) we have
\bea
   & & \Theta_{(2)}^{(0)}
       = {10 \over 9} {c_\tau \over H}
       \Bigg[ - \dot \Theta_{(2)}^{(0)}
       + {k \over a} \left( {1 \over 3} \kappa_2^0 v_{(\gamma)}
       - {1 \over 7} \kappa_3^0 \Theta_{(3)}^{(0)} \right)
   \nonumber \\
   & & \quad
       - {2 \over 3} \sqrt{ 1 - 3 {K \over k^2} }
       \left( {k \over a} \right)^2 \chi \Bigg] 
       - {\sqrt{6} \over 9} E_{(2)}^{(0)},
   \label{Theta-2-eq-tight}
\eea
where $c_\tau \equiv H/\dot \tau$.
Thus, $\Theta_{(2)}^{(0)}$ is of the $c_\tau$-
or $c_\tau {k \over aH}$-order
higher compared with $\delta_{(\gamma)}$, and we have
$\Theta_{(\ell+ 1)}^{(0)} \sim {k \over aH} c_\tau \Theta_{(\ell)}^{(0)}$.
{}From the $\ell =1$ component of eq. (\ref{Theta-lm-eq}) 
and eq. (\ref{v_b-eq-tight}) we have
\bea
   & & v_{(b)} - v_{(\gamma)} = {c_\tau \over H} {1 \over 1 + r}
       \Bigg[ - H v_{(b)} + \dot v_{(\gamma)} - \dot v_{(b)}
   \nonumber \\
   & & \quad
       + {k \over a} \left( c_{(b)}^2 \delta_{(b)}
       - {1 \over 4} \delta_{(\gamma)}
       + {1 \over 5} \kappa_2^0 \Theta^{(0)}_{(2)} \right) \Bigg].
   \label{v_b-v_p-eq}
\eea
Taking a time derivative of eq. (\ref{v_b-v_p-eq}) and using 
eqs. (\ref{v_b-eq-tight},\ref{v_b-v_p-eq}), 
to the first order in $c_\tau$ expansion, we can derive
\bea
   & & \dot v_{(b)} - \dot v_{(\gamma)} 
       = {2 r \over 1 + r} H \left( v_{(b)} - v_{(\gamma)} \right)
   \nonumber \\
   & & \quad
       - {c_\tau \over H} {1 \over 1 + r}
       \Bigg[ {6 r^2 \over (1 + r)^2} H^2
       \left( v_{(b)} - v_{(\gamma)} \right)
   \nonumber \\
   & & \quad
       + {k \over a} \left( H \alpha - c_{(b)}^2 \dot \delta_{(b)}
       + {1 \over 4} \dot \delta_{(\gamma)}
       + {1 \over 2} H \delta_{(\gamma)} \right) \Bigg],
   \label{v_b-v_gamma}
\eea
where we {\it assumed} the radiation era with $a \propto t^{1/2}$,
which applies {\it for Einstein gravity} with negligible $K$ and $\Lambda$
contributions in that era.
Thus, $\dot c_\tau = H c_\tau$, $\dot r = - H r$,
and we used $(c_{(b)}^{2})^\cdot = - H c_{(b)}^2$
which follows from eq. (\ref{c_b}) assuming $T_b = T_\gamma$.

To the zeroth order in $c_\tau$,
from eqs. (\ref{v_b-eq-tight},\ref{v_b-v_gamma}) we have:
\bea
   & & \dot v_{(b)} = \dot v_{(\gamma)}
       + {2 r \over 1 + r} H \left( v_{(b)} - v_{(\gamma)} \right)
   \nonumber \\
   & & \quad
   = - {1 \over 1 + r} H v_{(b)}
       - {2 r^2 \over (1 + r)^2 } H \left( v_{(\gamma)} - v_{(b)} \right)
   \nonumber \\
   & & \quad
       + {k \over a} \left( \alpha
       + {1 \over 1 + r} c_{(b)}^2 \dot \delta_{(b)}
       + {r \over 1 + r} {1 \over 4} \delta_{(\gamma)} \right).
\eea
Equations (\ref{v_b-eq},\ref{v_b-v_gamma}) imply 
$v_{(b)} - v_{(\gamma)} = 0$ to the zero-th order in $c_\tau$.
Thus, using $v_{\gamma b} \equiv v_{(b)} = v_{(\gamma)}$,
and ignoring the $c_{(b)}^2 \dot \delta_{(b)}$ term, we have
\bea
   & & \dot v_{\gamma b} 
       = - {1 \over 1 + r} H v_{\gamma b}
       + {k \over a} \left( \alpha
       + {r \over 1 + r} {1 \over 4} \delta_{(\gamma)} \right).
   \label{Tight-v-eq}
\eea
Equation (\ref{delta_b-eq}) and the $\ell = 0$ component of 
eq. (\ref{Theta-lm-eq}) give
\bea
   & & \dot \delta_{(b)} = - {k \over a} v_{\gamma b} - 3 H \alpha + \kappa,
   \label{Tight-delta_b} \\
   & & \dot \delta_{(\gamma)} = {4 \over 3} \left( 
       - {k \over a} v_{\gamma b} - 3 H \alpha + \kappa \right),
   \label{Tight-delta_p} 
\eea
and for the CDM we have eqs. (\ref{CDM-delta-eq},\ref{CDM-v-eq}).

Thus, in the tight coupling era, instead of
eqs. (\ref{Theta-lm-eq},\ref{v_b-eq}), we can use
and eqs. (\ref{Tight-v-eq}-\ref{Tight-delta_p}).
As the criteria for the tight coupling era we can use \cite{Bond-1996}
\bea
   & & z > z_{tc} = 2000, \quad
       c_\tau < 0.01, \quad
       c_\tau {k \over aH} < 0.01,
   \label{tight-conditions}
\eea
where
\bea
   & & c_\tau = {14.47 \over x_e \Omega_{b0} h}
       \left( {a \over a_0} \right)^3 {H \over H_0}.
\eea
If any one of the three conditions is violated we use the full set of
equations based on the Boltzmann equation.

\subsection{Numerical implementation}
                                                  \label{sec:Numerical}

The different gauge conditions available in the scalar-type perturbation 
provide a useful check of the numerical accuracy \cite{Viana-Liddle-1998}.
The gauge-ready formulation is especially suitable for handling the case.
When we solve the scalar-type perturbation equations we have a right to 
choose one temporal gauge condition.
Any one of the fundamental gauge conditions 
in eqs. (\ref{Gauges-1},\ref{Gauges-2}) would be a fine gauge condition; 
except for the synchronous gauge condition
any one of the other gauge conditions completely fixes the temporal gauge
and the remaining variables are equivalent to the gauge-invariant ones.
If we have the solution of a variable in a given gauge we can derive
solutions of the rest of variables in the same gauge, and from these
we can derive all the solutions in other gauge conditions.
{}For such translations the set of equations in the gauge-ready form
is convenient.
Meanwhile, in the numerical study, if we solve a given problem in two 
different gauge conditions independently, by comparing the value of 
any gauge-invariant variable evaluated in the two gauge conditions we can check 
the numerical accuracy.

In the literature, the synchronous gauge is the most widely adopted
gauge condition.
The synchronous gauge does not fix the gauge condition completely.
We can choose any of the gauge conditions mentioned in 
eqs. (\ref{Gauges-1},\ref{Gauges-2},\ref{GT-Theta},\ref{GT-hat-Theta}) as well.
The comoving gauge condition closely resembles the synchronous gauge
in the matter dominated era; however, see the various possible
combinations of the comoving gauge conditions in eqs. (\ref{Gauges-2}) 
available in the multi-component situation.
The authors of \cite{Lewis-etal-2000} adopted a comoving gauge condition 
which fixes the velocity variable based on the cold dark matter, 
thus $v_{(C)}/k \equiv 0$.
{}From eq. (\ref{CDM-v-eq}) we notice that $v_{(C)} = 0$ implies
$\alpha = 0$, the synchronous gauge condition;
in the synchronous gauge, however, we have $v_{(C)} \propto a^{-1}$
which leads to the remaining gauge mode.
Thus, the synchronous gauge with an additional condition $v_{(C)} = 0$ is
equivalent to the $v_{(C)} = 0$ gauge.
We may emphasize that our equations in the gauge-ready form are 
ready to be implemented using any of these available gauge conditions.

In the following, as an example, we consider {\it Einstein gravity limit}
without fields but with arbitrary numbers of fluids and kinetic components;
including the fields and the generalized gravity
will affect the gravity sector only.
In the numerical work we implemented the comoving gauge based on CDM 
($v_{(C)} = 0$ which includes the synchronous gauge), the zero-shear gauge, 
the uniform-curvature gauge and the uniform-expansion gauge.
The latter two gauge conditions were not previously used in the literature.
These four gauge conditions fix the perturbed metric variables.
Under these gauge conditions we can see that the differential equations
can be set up only using the variables which represent the fluids and 
kinetic components: eqs. 
(\ref{hat-Theta-lm-eq},\ref{Theta-lm-eq},\ref{E-eq},\ref{B-eq},
\ref{G9},\ref{G10})
provide a set of differential equations to be solved.
The remaining metric variables can be expressed in terms of 
the fluids and kinetic quantities.
The metric variables $\alpha$, $\varphi$, $H \chi$ (or $\chi/a$), 
and $\kappa/H$ (or $a \kappa$) are dimensionless.
Using eqs. (\ref{G1}-\ref{G5}) we can express the metric variables in terms 
of the fluid quantities in each of these gauge conditions.

In order to check the numerical accuracy we can evaluate any
gauge-invariant variable in two different gauge conditions;
if the integration has good numerical accuracy the gauge-invariant
combination evaluated in all gauge conditions should give the same value.
In order to have the same solution we need to start from the same initial
condition.
Thus, we need to have relations of variables among different gauge conditions.
The set of equations in a gauge-ready form is convenient for this purpose.
In the following we consider relations between the zero-shear gauge 
and the uniform-curvature gauge as an example.
Using the gauge transformation properties in 
eqs. (\ref{GT-1},\ref{GT-2},\ref{GT-i-1},\ref{GT-Theta},\ref{GT-hat-Theta}) 
we can construct the following relations
\bea
   & & \delta_{(i)\varphi} \equiv \delta_{(i)}
       + 3 ( 1 + w_{(i)} ) \varphi
       = \delta_{(i)\chi} + 3 ( 1 + w_{(i)} ) \varphi_\chi, 
   \nonumber \\
   & & v_{(i)\varphi} = v_{(i)\chi} - {k \over aH} \varphi_\chi, \quad
       \Theta^{(0)}_{(0)\varphi} = \Theta^{(0)}_{(0)\chi}
       + \varphi_\chi,
   \nonumber \\
   & & \Theta^{(0)}_{(1)\varphi} = \Theta^{(0)}_{(1)\chi}
       - {k \over aH} \varphi_\chi, 
   \label{UCG-ZSG}
\eea
and $\Theta^{(0)}_{(\ell)\varphi} = \Theta^{(0)}_{(\ell)\chi}$ for $\ell \ge 2$;
and similarly for $\hat \Theta^{(0)}_{(\ell)}$.
$\varphi_\chi$ follows from eqs. (\ref{G2},\ref{G3}) as
\bea
   \varphi_\chi 
   &=& {4 \pi G a^2 \over k^2 - 3K} \sum_l
       \left[ \delta \mu_{(l)\varphi}
       + 3 {a H \over k} ( \mu_{(l)} + p_{(l)} ) v_{(l)\varphi} \right] 
   \nonumber \\
   &=& - H \chi_\varphi.
   \label{varphi-chi}
\eea
Using eqs. (\ref{UCG-ZSG},\ref{varphi-chi}) we can translate the
solutions (including the initial conditions) in the zero-shear gauge
into the ones in the uniform-curvature gauge, and {\it vice versa}.

Each of the four gauge conditions considered above fixes the metric
variables as the gauge condition, and uses the fluid, field and the kinetic
variables as the unknown variables to be solved.
In such cases we can make the numerical code which allows to choose 
one of the gauge condition as an option.
In the numerical study it is known that the zero-shear gauge has 
numerical difficulty in setting up the initial condition at early universe 
\cite{Bardeen-1988,Ma-Bertschinger-1995}.
The other three gauge conditions show no such a difficulty encountered 
in the zero-shear gauge and run equally well.

We have implemented our gauge-ready formulation into the numerical code.
The code includes the baryon ($b$), CDM ($C$), photon ($\gamma$), 
massless ($\nu$) and massive neutrino ($\nu_m$)
species, the spatial curvature, and the cosmological constant.
We included photon polarizations.
We solved separately the gravitational wave with the accompanying tensor-type 
photon intensity and polarizations, massless and massive neutrino species. 
The set of differential equations is solved directly.
For the recombination process we adopt the RECfast code
by \cite{Seager-etal-1999} which is recently available in public.  
As the initial conditions, we have implemented the five different
non-decaying initial conditions available in the four component
($b$, $C$, $\gamma$ and $\nu$) system in the radiation dominated era
\cite{Bucher-etal-2000}: 
these are the adiabatic mode, the baryon isocurvature mode, 
the CDM isocurvature mode, the neutrino isocurvature density mode, 
and the neutrino isocurvature velocity mode, where the last one
appears due to the kinetic nature of the neutrino perturbation.
The complete set of initial conditions was recently presented by Bucher 
{\it et al} in \cite{Bucher-etal-2000} in the $v_{(C)} = 0$ gauge condition.
The corresponding initial conditions in the other gauge conditions
can be obtained by using the gauge transformations similar to
eqs. (\ref{UCG-ZSG},\ref{varphi-chi}).
In the early radiation dominated era we used the tight coupling
approximation for the baryon and the photon in 
eqs. (\ref{Tight-v-eq},\ref{Tight-delta_b},\ref{Tight-delta_p})
with the criteria in eq. (\ref{tight-conditions}).
The code is complete in the context of Einstein gravity.
We made no artifical truncations for multipoles of kinetic components
in the photon intensity and polarization, and massless/massive neutrinos.
{}For a useful truncation scheme, however, see \cite{Ma-Bertschinger-1995}.
Since the higher $\ell$ multipoles are generated from the lower multipoles, 
we monitered the values of highest multipoles of all the kinetic
components and increased the allowed multipoles automatically
\cite{Dodelson-etal-1996}.
In this manner we included the quantities with higher multipole as long as 
the values are larger than certain minimum threshold value.
We made the code so that we can choose the gauge condition out of the four
different gauge conditions mentioned above as an option; 
we could try other gauge conditions as well. 
The inclusion of the scalar fields and generalized gravity is 
a trivial generalization affecting only the gravity sector, 
and will be considered in future occasions.

At the present epoch we have
\bea
   & & {k \over a_0 H_0} = {2998 \over h} {k \over a_0} Mpc,
\eea
where $H_0 \equiv 100 h km/sec Mpc$.
Thus, $k/a_0 = 1Mpc^{-1}$ corresponds to $\lambda_0 = 2 \pi a_0/k = 2 \pi Mpc$,
and $k/(a_0 H_0) = 2998/h$; the comoving wave number $k$ is dimensionless.
In Einstein's gravity from eq. (\ref{BG1}) we have $K/(aH)^2 = \Omega -1$.
Thus, in the non-flat case we have
\bea
   & & a_0 = {2998 Mpc \over \sqrt{| 1 - \Omega_0 |} h }.
   \label{a_0}
\eea
In the hyperbolic model, the curvature scale corresponds to 
$k^2 = - K = 1$ and the subcurvature scales ($k^2 > 1$) correspond to
$k/a_0 > \sqrt{ 1- \Omega_0} h / (2998 Mpc)$.
In the spherical model, the wavenumber $n$ introduced in 
eq. (\ref{kappa-s-def}) takes integers $n = 3, 4, 5, \dots$
($n = 1, 2$ corresponds to pure gauge modes \cite{Lifshitz-1946}).

In the following we present several results from our numerical study.
In the numerical integration of the differential equations
we adopted the Runge Kutta method.
The integrations were made in the equal interval of $\ln{a}$.
In Figures \ref{fig:density1} (a,b)
we showed the evolution of $\delta_{(i)v_{(i)}}$ which is the
density perturbation of $(i)$-component
in the corresponding comoving gauge condition based on the component
\bea
   & & \delta_{(i)v_{(i)}} \equiv \delta_{(i)} + 3 {a H \over k}
       \left( 1 + w_{(i)} \right) v_{(i)},
\eea
which follows from eqs. (\ref{GT-i-1},\ref{BG5});
$w_{(i)} \equiv p_{(i)}/\mu_{(i)}$ and we ignored $Q_{(i)}$.
The component $(i)$ includes $(b)$, $(\gamma)$,
$(C)$, $(\nu)$ and $(\nu_m)$.
The small scale considered in Figure \ref{fig:density1} (a)
crosses the horizon in the radiation dominated era.
After the perturbation comes inside the horizon, the baryon, photon,
and massless neutrino show oscillations,
and after the recombination near $\log{(a/a_0)} \sim  -3$
the baryon decouples from the photon and catches up the
evolution of the  cold dark matter.
The behavior of massive neutrino is also shown.
The large scale perturbation in Figure \ref{fig:density1} (b)
crosses the horizon in the matter dominated era and
the oscillations do not appear.

In Figures \ref{fig:density1} (a,b)
we presented the behavior of $\varphi_v$ as well.
$\varphi_v$ was first introduced by Lukash in 1980 \cite{Lukash-1980}, 
see also \cite{Bardeen-1980}, and is known to be one of the best 
conserved quantities in the single component situation:
it is conserved independently of changing gravity theories or field potential
in the super-horizon scale \cite{Hwang-MSF-1994,Hwang-Noh-1996-GGT}, 
and independently of changing equation of state in the super-sound-horizon 
scale \cite{Hwang-Ideal-1993}; for a summary, 
see \cite{Noh-Hwang-2001-Unified}.
In the Figure it shows nearly conserved behavior while in the super-horizon 
scale and in the matter dominated era after the recombination;
however its amplitude changes near horizon crossing, and
is affected by the recombination process if the scale is inside horizon.
We showed detailed behaviors of $\varphi_v$ and $\varphi_{v_{(i)}}$
in Figures \ref{fig:varphi1} and \ref{fig:varphi2}.
{}Figure \ref{fig:varphi1} shows behaviors of $\varphi$ in
various comoving gauge conditions based on fixing $v$ or $v_{(i)}$
for three chosen scales.
We showed the evolution of $\varphi_v$ for several different scales
in Figure \ref{fig:varphi2}.

In Figures \ref{fig:massive3} (a,b)
we presented the evolution of $\delta_{(\nu_m) v_{(\nu_m)}}$ which is
the density perturbation of the massive neutrino in the comoving gauge
based on the massive neutrino.
In Figure \ref{fig:massive3} (a) we considered a model dominated by the
massive neutrino, showing the collisionless damping of the
neutrino density fluctautions \cite{Bond-Szalay-1983}.
The case with substantial amount of the cold dark matter is
presented in Figure \ref{fig:massive3} (b).
In the massive neutrino the fluid quantities include the integral of the
distribution function over the momentum variable, $q$ in 
eq. (\ref{fluid-massive}):
in our numerical work we considered about 100 values of $q$ 
for a range of $q/(k_B a_0 T_0)$.

As the wavenumber $k$ increases, i.e., as we consider the smaller scale,
we need to solve the larger number of the differential equations.
As examples, for $k/a_0 = 0.001$ and $0.1 Mpc^{-1}$ considered in
Figures \ref{fig:density1} (a,b)
$\ell$ is excited up to around 600 and 5000, respectively.
We increased $\ell$ automatically by monitoring the value of individual
kinetic component (including the polarizations).

Aspects of the roles of massless neutrino in the evolution of 
cosmic structures were studied in \cite{massless-nu}.
The roles of massive collisionless particles (massive neutrino is the
prime example) as the hot dark matter in the context of structure 
evolution have been investigated in the literature 
\cite{Peebles-1982,%
Bond-Szalay-1983,%
Bond-Efstathiou-1984,%
Durrer-1989,%
Ma-Bertschinger-1995,%
Bond-1996,%
Dodelson-etal-1996}.
{}Further analytic studies can be found in \cite{Zakharov-etal}.
The gravitational instability using the particle distribution function
was originally studied by Gilbert in 1965 in the Newtonian context
\cite{Gilbert-1965}.

\subsection{CMB anisotropy}
                                       \label{sec:CMB-Kinetic}

The anisotropies of the temperature can be derived by expanding the 
observed temperature in the sky into a spherical harmonic function as
\bea
   & & \Theta ({\bf x}, \eta_0, \hat {\bf \gamma})
       \equiv \sum_\ell \sum_{m = -\ell}^\ell a^\Theta_{\ell m} ({\bf x})
       Y_{\ell m} (\hat {\bf \gamma}).
\eea
The polarization anisotropies can be expanded in terms of the
spin-weighted harmonic functions as 
\bea
   & & Q ({\bf x}, \eta_0, \hat {\bf \gamma})
       \pm i U ({\bf x}, \eta_0, \hat {\bf \gamma})
   \nonumber \\
   & & \quad
       \equiv \sum_\ell \sum_{m = -\ell}^\ell \left[
       a^E_{\ell m} ({\bf x}) \pm i a^B_{\ell m} ({\bf x}) \right]
       {}_{\pm 2} Y_{\ell m} (\hat {\bf \gamma}).
\eea
We can derive:
\bea
   & & C_\ell^{XY} \equiv {1 \over 2 \ell + 1} \sum_{m = -\ell}^\ell
       \langle a^X_{\ell m} ({\bf x}) a^{Y*}_{\ell m} ({\bf x}) \rangle_{\bf x}
   \nonumber \\
   & & \quad
       = {1 \over (2 \ell + 1)^2} {2 \over \pi}
       \int n^2 d n \sum_{m = -2}^2 X_{(\ell)}^{(m)} (n, t_0)
       Y_{(\ell)}^{(m)*} (n, t_0),
   \nonumber \\
   \label{C_ell-def}
\eea
where $X$ and $Y$ can be any one of $\Theta$, $E$ and $B$.
In the flat and hyperbolic background ($K \le 0$) we have $n \ge 0$,
see below eq. (\ref{kappa-s-def}).
In the background with positive curvature, we have discrete $n$
with $n = 3, 4, \dots$; see below eq. (\ref{a_0}). 
In such a case the integration should be changed to a sum over
$n$ with $n = 3, 4, \dots$.
Due to the parity, we have $C_\ell^{\Theta B} = 0 = C_\ell^{EB}$
\cite{Hu-etal-1998}.
If the distributions are Gaussian, all statistical informations are
contained in the three angular power spectra and one correlation
power spectrum between $\Theta$ and $E$:
\bea
   & & C_\ell^{\Theta\Theta}, \quad C_\ell^{EE}, \quad C_\ell^{BB},
       \quad C_\ell^{\Theta E}.
   \label{C_ell}
\eea
Both the scalar-type and gravitational wave contribute to the correlation 
functions $C_\ell^{\Theta\Theta}$, $C_\ell^{EE}$ and $C_\ell^{\Theta E}$, 
whereas only the gravitational wave (and the rotation) 
contributes to $C_\ell^{BB}$ \cite{polarization-GW}.

In Figures \ref{fig:Cl} (a,b) and Figure \ref{fig:ClGW} we presented 
the power spectra of the scalar, and tensor-type perturbations.
In both the scalar- and tensor-type perturbation spectra,
for the integration over $k$ in eq. (\ref{C_ell-def}),
we took 500 $k$'s in the equal interval of $\log{k}$ for a range
from $k/a_0=10^{-4} \sim 0.5 Mpc^{-1}$;
to have $C_\ell$ complete to $\ell \sim 2000$ (which would well
cover the MAP and Planck Surveyor results) we need
$k_{max}/a_0 \sim 0.2 Mpc^{-1}$.
The spectra are filtered using a smoothing method.

Pioneering work concerning the CMB anisotropy based on relativistic
gravity and Boltzmann equation was made by Peebles and Yu in 1970
\cite{Peebles-Yu-1970}.
Early theoretical works can be found in 
\cite{Doroshkevich-etal-1978,%
Bond-Efstathiou-1987,%
Pre-COBE}.
Significant progress has been made in the theoretical side of the CMB 
anisotropy immediately following the first detection of the quadrupole
and higher-order multipole anisotropies by the {\it COBE}-DMR and 
subsequent ground based experiments; some selected references are 
\cite{Post-COBE,%
Bond-1996,%
polarization-GW,%
polarization-others,%
Ma-Bertschinger-1995,%
Schaefer-deLaix-1996,%
Seager-etal-1999,%
Hu-etal-1998,%
covariant-pert}.
Authors of \cite{Seljak-Zaldarriaga-1996} have developed a CMBfast code 
which calculates the CMB angular power spectra in an efficient way
using the line of sight integration of the Boltzmann's equation.
CMBfast is based on the synchronous gauge and is applicable to Einstein gravity.
Authors of \cite{Lewis-etal-2000} have modified the code adopting a comoving 
gauge condition based on the velocity variable of the cold dark matter
(our $v_{(C)}/k = 0$ gauge).   
In our code, we solve the full Boltzmann hierarchy without any approximation.
A run takes less than an hour (without the massive neutrino)
in the workstation equally well for all three gauge conditions we used.

The power spectra in eq. (\ref{C_ell}) are known to be sensitive to 
various combinations of the background world models (these include
the Hubble constant, spatial curvature, cosmological constant,
and density parameters of baryon, CDM, massless and massive neutrinos), 
the initial amplitudes and spectra of both the primordial density and
gravitational wave, and the possible reionization history, etc. 
Thus, in return, the observational progress in determining the
power spectra can give strong constraints on the above mentioned
parameters with higher precisions \cite{Bond-etal-1997}.

There has been a significant improvement of the CMB power spectrum 
measurements in the past decade, and further improvements are 
expected from the ground based, balloon, and flight experiments, 
and particularly from the planned MAP and Planck Surveyor satellite 
missions with a high-accuracy and small angular resolution.
The recent balloon observations of CMB by Boomerang and Maxima-1 experiments
have already provided a strong constraint on the global curvature of 
our observed patch of the universe: the location of first peak in 
Figures \ref{fig:Cl} (a,b) corresponds, in models with $\Omega_0$ near $1$, to 
$\ell_{\rm peak} \simeq 200/\sqrt{\Omega_0}$ whereas the Boomerang experiment
shows $\ell_{\rm peak} = 197 \pm 6$, thus supporting a flat universe
\cite{deBernardis-etal-2000}.

The small $\ell$ plateau region in Fig. \ref{fig:Cl} can be
interpreted as reflecting the primordial scale-invariant spectrum,
which has arisen from the quantum fluctuations 
in the context of the inflationary scenario.
The small $\ell$ corresponds to the large angular scale, and the plateau
region corresponds to the super-horizon scale in the last scattering epoch
where the local scattering would be unimportant.
Thus, the Sachs-Wolfe effect based on the null-geodesic equations
is expected to be enough to explain the physics (based on the 
relativistic gravitation relating the spacetime metric with matter).
Meanwhile, the oscillatory features in the large $\ell$
(small angular scale) come from regions well inside the horizon
at the last scattering, thus in addition to the gravity the local physics 
including direct couplings between photons and baryons is important.
Now, the physics behind these oscillatory features is well 
understood as being due to the oscillatory evolution of the photon
fluctuation (and tightly coupled baryons) pictured/frozen at the
last scattering epoch: the oscillatory evolution of each $k$ mode
(reaching the last scattering epoch with different phase)
together with the initial spectrum is reflected into the 
corresponding oscillatory feature in $k$ space which can be converted
into the oscillatory feature of $C_\ell$ in the $\ell$ (angular) space.
In hindsight, the original prediction of this oscillatory
feature can be traced back to Sakharov as early as 1966 \cite{Sakharov-1966}
(which was before the discovery of CMB); for a clear exposition, 
see \cite{Sunyaev-Zeldovich-1970,Zeldovich-Novikov-1983,Albrecht-1996},
and for more elaborated forms see \cite{Post-COBE}.
More visually, the oscillatory feature in $k$ space can be found in 
Peebles and Yu in 1970 \cite{Peebles-Yu-1970} 
and Sunyaev and Zeldovich in 1970 \cite{Sunyaev-Zeldovich-1970}, 
and more developed analysis was made in Doroshkevich {\it et al} 
in 1978 \cite{Doroshkevich-etal-1978}.
Up to our knowledge, however, the first clear and complete picture of
the oscillatory features (including the polarization as well
as the isocurvature case) in Fig. \ref{fig:Cl} can be found in 
Figure 7 of Bond and Efstathiou (1987) \cite{Bond-Efstathiou-1987}.

\section{Discussions}
                                       \label{sec:Discussion}

Compared with the previous works we have made some
notable advances in the formulation.
The formulation is made for the general form of the Lagrangian in 
eq. (\ref{Lagrangian}) which is more general compared with our previous works.
The kinetic theory treatments in \S \ref{sec:Kinetic} and 
\ref{sec:CMB} are presented in a gauge-ready form for the scalar-type
perturbation.
Also, the kinetic theory formulation is made in the full context of
the generalized gravity theory covered by the Lagrangian in 
eq. (\ref{Lagrangian}).

{}For the scalar-type structure all the equations are arranged in a
gauge-ready form which enables the optimal use of various gauge conditions
depending on the problem.
Usually we do not know the most suitable gauge condition {\it a priori}.
In order to take the advantage of gauge choice in the most optimal way
it is desirable to use the gauge-ready form equations presented in this paper.
Our set of equations is arranged so that we can easily impose various 
fundamental gauge conditions in eqs. (\ref{Gauges-1},\ref{Gauges-2}),
and their suitable combinations as well.
Our notation for gauge-invariant combinations proposed in eq. (\ref{GI-single})
is practically convenient for connecting solutions in different gauge
conditions as well as tracing the associated gauge conditions easily.

In handling the Boltzmann equations numerically, we showed
uniform-expansion gauge and the uniform-curvature gauge
could also handle the numerical integration successfully.
By comparing solutions solved separately in different 
gauge conditions we can naturally check the numerical accuracy.
It may deserve examining the physics of CMB temperature and polarization
anisotropies in the persepective of these new gauge conditions and
others which might still deserve closer look. 
Our set of equations in a gauge-ready form is particularly suitable for 
such investigations where we can easily switch our perspective based on
one gauge condition to the other.

In this paper one can find the general cosmological perturbation equations 
which are ready for use in diverse FLRW world models based on the gravity 
theories in eq. (\ref{Lagrangian}).
More attention will be paid on the generalized versions of
gravity theories especially in the context of early universe in future.
In such a context, the formulation made in this
paper would be useful for studying the structure formation 
aspects of the future cosmological models.

\section*{Acknowledgments}
\addcontentsline{toc}{section}{Acknowledgements}

We thank Heewon Lee and Antony Lewis for useful discussions,
and Oystein Elgaroy for help in numerical study.
We also wish to thank George Efstathiou and Ofer Lahav 
for their encouragements and useful discussions.
HN was supported by grant No. 2000-0-113-001-3 from the
Basic Research Program of the Korea Science and Engineering Foundation.
JH was supported by Korea Research Foundation Grants (KRF-99-015-DP0443
and 2000-015-DP0080).

\appendix
\section{Conformal transformation}
                                      \label{sec:CT}
\setcounter{equation}{0}
\def\theequation{A\arabic{equation}}

By the conformal transformation the gravity theories included 
in eq. (\ref{Lagrangian}) can be transformed into Einstein gravity \cite{CT}.
Under the conformal transformation of the spacetime metric,
$\hat g_{ab} = \Omega^2 g_{ab}$, and the field redefinition,
$\Omega \equiv \sqrt{8 \pi G F} \equiv e^{{1 \over 2}\sqrt{2 \over 3} \psi}$,
eq. (\ref{Lagrangian}) becomes 
\bea
   & & \hat L = {1 \over 16 \pi G} \hat R
       - {1\over 16 \pi G} \left( \psi^{\hat ; c} \psi_{,c}
       + {1 \over F} g_{IJ} \phi^{I \hat ;c} \phi^J_{\;\;,c} \right) - \hat V,
   \nonumber \\
   \label{Lagrangian-CT1}
\eea
with
\bea
   & & \hat V \equiv {1 \over (16 \pi G F)^2} \left( 2 V - f + RF \right),
\eea
where we have ignored $L_m$ term.
Thus, in general, since $\psi = \psi (\phi^K, R)$,
we have an additional minimally coupled scalar field $\psi$.
However, if $\psi = \psi(\phi^K)$
which is the case for $f = F(\phi^K)R$ and for the gravity theories
in eq. (\ref{gravities}), eq. (\ref{Lagrangian-CT1}) becomes 
\bea
   & & \hat L = {1 \over 16 \pi G} \hat R
       - {1 \over 2} \hat g_{IJ} \phi^{I \hat ;c} \phi^J_{\;\;,c} - \hat V,
   \label{Lagrangian-CT2}
\eea
where
\bea
   & & \hat g_{IJ} \equiv {1 \over 8 \pi G} \left( {1 \over F} g_{IJ} 
       + \psi_{,I} \psi_{,J} \right). 
\eea
Relations we need to derive the above results can be found in
\cite{Hwang-GGT-1990,Hwang-GGT-CT-1997}.

As a simpler situation we consider a case with $g_{1\ell} =0$ 
and $\psi = \psi(\phi)$
where $\ell, m = 2, 3, \dots, N$, and $\phi \equiv \phi^1$.
By introducing
\bea
   & & d \hat \phi \equiv \sqrt{ {1 \over 8 \pi G} \left( 
       {g_{11} \over F} d \phi^2 + d \psi^2 \right) },
\eea
we can show that eq. (\ref{Lagrangian-CT1}) becomes
\bea
   & & \hat L = {1 \over 16 \pi G} \hat R
       - {1\over 2} \left( \hat \phi^{\hat ; c} \hat \phi_{,c}
       + {1 \over 8 \pi G F} g_{\ell m} \phi^{\ell \hat ;c} 
       \phi^m_{\;\;,c} \right) - \hat V,
   \nonumber \\
   \label{Lagrangian-CT3}
\eea
where we have a canonical form kinetic term of $\hat \phi$.
Equation (\ref{Lagrangian-CT3}) also follows directly from 
eq. (\ref{Lagrangian-CT2}).
Notice that eqs. 
(\ref{Lagrangian-CT1},\ref{Lagrangian-CT2},\ref{Lagrangian-CT3})
all belong to our original Lagrangian in eq. (\ref{Lagrangian}).

The conformal transformation in the context of cosmological perturbation
has been considered in 
\cite{Salopek-etal-1989,Hwang-GGT-1990,Hwang-GGT-CT-1997}.
We decompose the conformal factor $\Omega$ into the background and
the perturbed part as
\bea
   & & \Omega ({\bf x}, t) \equiv \bar \Omega (t)
       \Big[ 1 + \delta \Omega ({\bf x}, t) \Big].
   \label{Omega-pert}
\eea
Thus, we have:
\bea
   & & \bar \Omega = \sqrt{8 \pi G \bar F} 
       = e^{{1\over 2} \sqrt{{2\over 3}} \bar \psi}, \quad 
       \delta \Omega = {\delta F \over 2 F}
       = {1\over 2} \sqrt{2\over 3} \delta \psi.
   \label{Omega-psi}
\eea
In \cite{Hwang-GGT-1990,Hwang-GGT-CT-1997} we have shown that the only 
changes under the conformal transformation are the following 
\bea
   & & \hat a = a \bar \Omega, \quad d \hat t = \bar \Omega d t, \quad
       \hat \alpha = \alpha + \delta \Omega, \quad
       \hat \varphi = \varphi + \delta \Omega.
   \label{CT-pert1}
\eea
Thus, in our multicomponent situation, assuming the conditions used
to derive eq. (\ref{Lagrangian-CT3}) are met, we have:
\bea
   & & \hat H = {1\over \Omega} \left( H + {\dot \Omega \over \Omega} \right),
       \quad \hat \chi = \Omega \chi, 
   \nonumber \\
   & & \dot {\hat \phi} = \sqrt{ {1 \over 8 \pi G} 
       \left( {g_{11} \over F} \dot \phi^2
       + {3 \dot F^2 \over 2 F^2} \right)}, \quad
       {\delta \hat \phi \over \dot {\hat \phi}}
       = {\delta \phi \over \dot \phi}
       = {\delta F \over \dot F}.
   \label{CT-pert2}
\eea
[In \cite{Hwang-GGT-1990,Hwang-GGT-CT-1997} we considered the situation 
with a single field with $g_{11} = \omega(\phi)$.
In the present case $g_{11}$ and $g_{\ell m}$ are arbitrary algebraic
functions of $\phi$ and $\phi^\ell$.]
{}From these we can also show that
\bea
   & & d \eta, \quad \nabla^2, \quad k, \quad 
       \varphi_{\delta \phi} = - {H \over \dot \phi} \delta \phi_\varphi, \quad 
       C_{\alpha\beta}^{(t)},
   \label{CT-pert3}
\eea
are invariant under the conformal transformation.
Relations among $\hat \phi$, $\phi$, and $F$ in the individual
gravity are summarized in Table 2 of \cite{Hwang-GGT-CT-1997}.
The advantages of using the conformal transformation in cosmological
perturbation as a mathematical trick to simplify the analyses
are presented in \cite{Hwang-GGT-1990,Hwang-GGT-CT-1997}.

\section{Effective fluid quantities}
                                         \label{sec:Fluid-quantities}

\setcounter{equation}{0}
\def\theequation{B\arabic{equation}}

We present the effective fluid quantities based on the effective
energy-momentum tensor introduced in eq. (\ref{GFE}).
The effective energy-momentum tensor in eq. (\ref{GFE}) is decomposed
into the effective fluid quantities similarly as in eqs. 
(\ref{Tab-f},\ref{Tab-decomp},\ref{Tab}).
To the background order we have:
\bea
   & & 8 \pi G \mu^{({\rm eff})}
       = {1 \over F} \Big[ \mu + {1 \over 2} g_{IJ} \dot \phi^I \dot \phi^J
       - {1 \over 2} \left( f - RF - 2 V \right)
   \nonumber \\
   & & \quad
       - 3 H \dot F \Big],
   \nonumber \\
   & & 8 \pi G p^{({\rm eff})}
       = {1 \over F} \Big[ p + {1 \over 2} g_{IJ} \dot \phi^I \dot \phi^J
       + {1 \over 2} \left( f - RF - 2 V \right)
   \nonumber \\
   & & \quad
       + \ddot F + 2 H \dot F \Big],
   \nonumber \\
   & & q_a^{({\rm eff})} = 0 = \pi_{ab}^{({\rm eff})}.
   \label{eff-BG}
\eea
The scalar-type perturbed order effective fluid quantities are
[use eqs. (B4,B5) in \cite{Hwang-Noh-1996-GGT}]:
\bea
   & & 8 \pi G \delta \mu^{({\rm eff})} = {1 \over F} \Bigg\{ \delta \mu
       + g_{IJ} \dot \phi^I \delta \dot \phi^J
   \nonumber \\
   & & \quad
       + {1 \over 2} \left[ g_{IJ,K} \dot \phi^J \dot \phi^J
       - \left( f - 2 V \right)_{,K} \right] \delta \phi^K
   \nonumber \\
   & & \quad
       - 3 H \delta \dot F
       + \left( 3 \dot H + 3 H^2 - {k^2 \over a^2} \right) \delta F
   \nonumber \\
   & & \quad
       + \dot F \kappa
       + \left( 3 H \dot F - g_{IJ} \dot \phi^I \dot \phi^J \right) \alpha
       \Bigg\},
   \nonumber \\
   & & 8 \pi G \delta p^{({\rm eff})} = {1 \over F} \Bigg\{ \delta p
       + g_{IJ} \dot \phi^I \delta \dot \phi^J
   \nonumber \\
   & & \quad
       + {1 \over 2} \left[ g_{IJ,K} \dot \phi^I \dot \phi^J
       + \left( f - 2 V \right)_{,K} \right] \delta \phi^K
   \nonumber \\
   & & \quad
       + \delta \ddot F + 2 H \delta \dot F
       - \left( \dot H + 3 H^2 - {2\over 3} {k^2 - 3K \over a^2} \right)
       \delta F
   \nonumber \\
   & & \quad
       - \dot F \left( \dot \alpha + {2 \over 3} \kappa \right)
       - \left( 2 \ddot F + 2 H \dot F + g_{IJ} \dot \phi^I \dot \phi^J
       \right) \alpha \Bigg\},
   \nonumber \\
   & & 8 \pi G (\mu^{({\rm eff})} + p^{({\rm eff})}) v^{({\rm eff})}
       = {1 \over F} ( \mu + p ) v
   \nonumber \\
   & & \quad
       + {k \over a} {1 \over F} \left( g_{IJ} \dot \phi^I \delta \phi^J
       + \delta \dot F - H \delta F - \dot F \alpha \right),
   \nonumber \\
   & & 8 \pi G \pi^{(s,{\rm eff})}
       = {1\over F} \left[ \pi^{(s)} + {k^2 \over a^2}
       ( \delta F - \dot F \chi ) \right].
   \label{eff-pert}
\eea

The vector-type effective energy-momentum tensor is:
\bea
   & & 8 \pi G \delta T^{(v,{\rm eff})0}_{\;\;\;\;\;\;\;\;\;\;\alpha}
       = {1 \over F} \delta T^{(v)0}_{\;\;\;\;\;\alpha},
   \nonumber \\
   & & 8 \pi G \delta T^{(v,{\rm eff})\alpha}_{\;\;\;\;\;\;\;\;\;\;\beta}
       = {1 \over F} \delta T^{(v)\alpha}_{\;\;\;\;\;\beta}
   \nonumber \\
   & & \quad
       - {\dot F \over F} {1 \over 2 a} \left[ B^\alpha_{\;\;|\beta}
       + B_\beta^{\;\;|\alpha} + a \left( C^\alpha_{\;\;|\beta}
       + C_\beta^{\;\;|\alpha} \right)^\cdot \right].
   \label{eff-rot}
\eea

The tensor-type effective energy-momentum tensor is:
\bea
   & & 8 \pi G \delta T^{(t,{\rm eff})\alpha}_{\;\;\;\;\;\;\;\;\;\;\beta}
       = {1 \over F} \left( \delta T^{(t)\alpha}_{\;\;\;\;\;\beta}
       - \dot F \dot C^\alpha_\beta \right).
   \label{eff-GW}
\eea

\section{Kinematic quantities}
                                                 \label{sec:Kinematic}
\setcounter{equation}{0}
\def\theequation{C\arabic{equation}}

The $3+1$ ADM equations \cite{ADM}
and the $1+3$ covariant equations \cite{covariant}
are convenient in analysing the cosmological perturbations, 
\cite{Bardeen-1980,Hawking-1966,Ellis-Bruni-1989,Hwang-Vishniac-1990}.
The kinematic quantities and the Weyl curvatures appearing in the
formulations are useful to characterize the variables used
in the perturbation analyses.
In the following we present various quantities appearing in the
two formulations in the context of our perturbed FLRW metric.
{}For the basic sets of the ADM and the covariant equations,
see \S VI in \cite{Bardeen-1980}, \cite{covariant} and
the Appendix in \cite{Hwang-Vishniac-1990}.

The covariant decomposition of the normalized ($n^a n_a \equiv -1$)
normal-frame vector field $n_a$ provides
clear meanings of the perturbed metric variables.
The normal-frame vector field is introduced as
\bea
   & & n_0 \equiv - a ( 1 + A ), \quad
       n_\alpha \equiv 0.
   \label{n_a}
\eea
The kinematic quantities based on the normal-frame vector are
\bea
   & & \hat \theta_{ab} \equiv \hat h_a^c \hat h_b^d n_{(c;d)}
       = n_{(a;b)} + \hat a_{(a} n_{b)}, \quad
       \hat \theta \equiv n^a_{\;\;;a},
   \nonumber \\
   & & \hat \sigma_{ab} \equiv \hat \theta_{ab} - {1\over 3} \hat \theta
       \hat h_{ab}, \quad
       \hat a_a \equiv n_{a;b} n^b,
   \label{kinematic-n-def}
\eea
where $t_{(ab)} \equiv {1 \over 2} ( t_{ab} + t_{ba} )$
and $t_{[ab]} \equiv {1 \over 2} ( t_{ab} - t_{ba} )$.
$\hat h_{ab} \equiv g_{ab} + n_a n_b$ is the projection tensor based on $n_a$.
$\hat \theta$, $\hat \sigma_{ab}$, and $\hat a_a$ are the
expansion scalar, shear tensor, and the acceleration vector
based on $n_a$, respectively.
The vorticity tensor of the normal-vector, $\hat \omega_{ab}$
naturally vanishes, see eq. (\ref{kinematic-tilde-u-def}).
{}From eq. (\ref{kinematic-n-def}) using
eqs. (\ref{metric-general},\ref{n_a},\ref{metric-decomp},\ref{chi-kappa})
we can show
\bea
   & & \hat \theta = 3 H - \kappa,
   \nonumber \\
   & & \hat \sigma_{\alpha\beta} = \chi_{,\alpha|\beta}
       - {1\over 3} g_{\alpha\beta}^{(3)} \Delta \chi
       + a \Psi^{(v)} Y^{(v)}_{(\alpha|\beta)}
       + a^2 \dot c^{(t)} Y^{(t)}_{\alpha\beta},
   \nonumber \\
   & & \hat a_\alpha = \alpha_{,\alpha}.
   \label{kinematic-n}
\eea
Therefore, $-\kappa$ and $\chi$ can be interpreted as the perturbed
expansion scalar and
the scalar part of shear of the normal-frame, respectively.
The trace of the extrinsic curvature is equal to minus of the expansion scalar.
$\alpha$ and $\beta$ can be seen as perturbations in the lapse function and
shift vector, respectively.
$\Psi^{(v)}$ and $\dot c^{(t)}$ also cause the shear in the perturbed
normal hypersurface.

In order to interprete the velocity related quantities we introduce
frame-invariant combinations of the four-vectors as \cite{Israel-1976}
\bea
   & & \tilde u_a \equiv u_a + {q_a \over \mu + p}.
   \label{tilde-u-def}
\eea
Similarly as in eq. (\ref{kinematic-n-def}) we can introduce the kinematic
quantities based on $\tilde u_a$ vector
\bea
   & & \theta_{ab} \equiv \tilde h_a^c \tilde h_b^d u_{(c;d)}
       = \tilde u_{(a;b)} + a_{(a} \tilde u_{b)}, \quad
       \theta \equiv {\tilde u}^a_{\;\;;a}, \quad
   \nonumber \\
   & & \sigma_{ab} \equiv \theta_{ab} - {1\over 3} \theta \tilde h_{ab}, \quad
   \nonumber \\
   & & \omega_{ab} \equiv \tilde h_a^c \tilde h_b^d u_{[c;d]}, \quad
       a_a \equiv \tilde u_{a;b} \tilde u^b,
   \label{kinematic-tilde-u-def}
\eea
where $\tilde h_{ab} \equiv g_{ab} + \tilde u_a \tilde u_b$
is the projection tensor based on $\tilde u_a$.
$\theta$, $\sigma_{ab}$, $\omega_{ab}$, and $a_a$ are the
expansion scalar, shear tensor, vorticity tensor, and the acceleration vector
based on $\tilde u_a$, respectively.
{}From eq. (\ref{kinematic-tilde-u-def}) using
eqs. (\ref{tilde-u-def},\ref{metric-general},\ref{metric-decomp},\ref{chi-kappa}
,\ref{fluid-decomp})
we can show
\bea
   & & \tilde u_0 = u_0, \quad
       \tilde u_\alpha = - {a \over k} v^{(s)}_{,\alpha}
       + a v^{(v)} Y_\alpha^{(v)},
   \nonumber \\
   & & \theta = 3 H - \kappa + {k \over a} v^{(s)},
   \nonumber \\
   & & \sigma_{\alpha\beta} = \chi_{,\alpha|\beta} - {1\over 3}
       g_{\alpha\beta}^{(3)} \Delta \chi
       - {a \over k} \left( v^{(s)}_{,\alpha|\beta}
       - {1 \over 3} \Delta v^{(s)} g^{(3)}_{\alpha\beta} \right)
   \nonumber \\
   & & \quad
       + a \left( \Psi^{(v)} + v^{(v)} \right) Y^{(v)}_{(\alpha|\beta)}
       + a^2 \dot c^{(t)} Y^{(t)}_{\alpha\beta},
   \nonumber \\
   & & \omega_{\alpha\beta} = a v^{(v)} Y^{(v)}_{[\alpha|\beta]},
   \nonumber \\
   & & a_\alpha = \left[ \alpha - {1 \over k} ( a v^{(s)} )^\cdot
       \right]_{,\alpha} + ( a v^{(v)} )^\cdot Y_\alpha^{(v)},
   \label{kinematic-tilde-u}
\eea
and similarly for the kinematic quantities based on the individual
fluid four-vectors $\tilde u_{(i)a}$.

Weyl curvature tensor is introduced as
\bea
   & & C_{abcd} \equiv R_{abcd} - {1 \over 2} \left(
       g_{ac} R_{bd} + g_{bd} R_{ac} - g_{bc} R_{ad} - g_{ad} R_{bc} \right)
   \nonumber \\
   & & \quad
       + {R \over 6} \left( g_{ac} g_{bd} - g_{ad} g_{bc} \right).
   \label{Weyl-tensor}
\eea
It is decomposed into the electric and the magnetic parts as
\bea
   & & E_{ab} \equiv C_{acbd} u^c u^d, \quad
       H_{ab} \equiv {1 \over 2} \eta_{ac}^{\;\;\;\;ef} C_{efbd} u^c u^d.
\eea
Both are symmetric, tracefree and orthogonal to $u^a$;
$E_{ab} = E_{ba}$, $E^a_a = 0 = E_{ab} u^b$, and the same for $H_{ab}$.
The nonvanishing electric and magnetic parts of the Weyl curvature are:
\bea
   E_{\alpha\beta}
   &=& - C^0_{\;\;\;\alpha 0\beta}
   \nonumber \\
   &=& {1 \over 2} k^2 \left( \alpha - \varphi - \dot \chi + H \chi
       \right) Y^{(s)}_{\alpha\beta}
       + {1 \over 2} a k \dot \Psi^{(v)} Y^{(v)}_{\alpha\beta}
   \nonumber \\
   & & - {1 \over 2} a^2 \left[ \ddot c^{(t)} + H \dot c^{(t)}
       + {\Delta - 2 K \over a^2} c^{(t)} \right] Y^{(t)}_{\alpha\beta},
   \nonumber \\
   H_{\alpha\beta}
   &=& - {1 \over 2} \eta_{0(\alpha}^{\;\;\;\;\;\;\gamma\delta}
       C^0_{\;\;\;\beta)\gamma\delta}
   \nonumber \\
   &=& \eta_{0(\alpha}^{\;\;\;\;\;\;\gamma\delta}
       \left( - k  \Psi^{(v)} Y^{(v)}_{\beta)\gamma|\delta}
       + a \dot c^{(t)} Y^{(t)}_{\beta)\gamma|\delta} \right),
\eea
which follow from the Riemann curvature tensors
and eqs. (\ref{Y-def},\ref{chi-kappa}).

In the ADM notation
\bea
   & & g_{00} \equiv - N^2 + N^\alpha N_\alpha, \quad
       g_{0\alpha} \equiv N_\alpha, \quad
       g_{\alpha\beta} \equiv h_{\alpha\beta},
\eea
where $N_\alpha$ is based on $h_{\alpha\beta}$
with the $h^{\alpha\beta}$ the inverse metric; only in the rest of this
Appendix $h_{\alpha\beta}$ indicates the ADM three-space metric.
The normal four-vector is $n_0 \equiv - N$ and $n_\alpha \equiv 0$.
The extrinsic curvature is
\bea
   & & K_{\alpha\beta} \equiv {1 \over 2N} \left( N_{\alpha:\beta}
       + N_{\beta:\alpha} - h_{\alpha\beta,0} \right),
   \nonumber \\
   & & K \equiv h^{\alpha\beta} K_{\alpha\beta},
\eea
where a colon `$:$' indicates a covariant derivative based on $h_{\alpha\beta}$.
$\Gamma^{(h)\alpha}_{\;\;\;\;\;\beta\gamma}$ is the connection based on
$h_{\alpha\beta}$.
The ADM fluid quantities are
\bea
   & & E \equiv n_a n_b T^{ab}, \quad
       J_\alpha \equiv - n_b T^b_\alpha, \quad
       S_{\alpha\beta} \equiv T_{\alpha\beta},
   \nonumber \\
   & & S \equiv h^{\alpha\beta} S_{\alpha\beta}, \quad
       \bar S_{\alpha\beta} \equiv S_{\alpha\beta}
       - {1 \over 3} h_{\alpha\beta} S.
\eea
Compared with the perturbed metric in eq. (\ref{metric-general})
we have
\bea
   & & h_{\alpha\beta} = a^2 \left( g^{(3)}_{\alpha\beta}
       + 2 C_{\alpha\beta} \right),
   \nonumber \\
   & & N = a ( 1 + A ), \quad
       N_\alpha = - a^2 B_\alpha, \quad
   \nonumber \\
   & & \Gamma^{(h)\alpha}_{\;\;\;\;\;\beta\gamma}
       = \Gamma^{(3)\alpha}_{\;\;\;\;\;\beta\gamma}
       + C^\alpha_{\beta|\gamma} + C^\alpha_{\gamma|\beta}
       - C_{\beta\gamma}^{\;\;\;\;|\alpha}.
\eea
Thus we can show
\bea
   & & K_{\alpha\beta} = - a \left[ {a^\prime \over a} g^{(3)}_{\alpha\beta}
       ( 1 - A ) + B_{(\alpha|\beta)}
       + C^\prime_{\alpha\beta} + 2 {a^\prime \over a} C_{\alpha\beta}
       \right],
   \nonumber \\
   & & K = - {1 \over a} \left[ 3 {a^\prime \over a} ( 1 - A )
       + B^\alpha_{\;\;|\alpha} + C^{\alpha\prime}_\alpha \right]
       = - 3 H + \kappa
       = - \hat \theta,
   \nonumber \\
   & & R^{(h)\alpha}_{\;\;\;\;\;\;\;\beta\gamma\delta}
       = R^{(3)\alpha}_{\;\;\;\;\;\;\;\beta\gamma\delta}
       + C^\alpha_{\beta|\delta\gamma} - C^\alpha_{\beta|\gamma\delta}
       + C^\alpha_{\delta|\beta\gamma} - C^\alpha_{\gamma|\beta\delta}
   \nonumber \\
   & & \quad
       - C_{\beta\delta\;\;\;\gamma}^{\;\;\;\;|\alpha}
       + C_{\beta\gamma\;\;\;\delta}^{\;\;\;\;|\alpha},
   \nonumber \\
   & & R^{(h)} = {1 \over a^2} \left[ 6 \bar K 
       - 4 ( \Delta + 3 \bar K ) \varphi \right],
   \nonumber \\
   & & E = -T^0_0 = \mu, \quad
       J_\alpha = a T^0_\alpha = q_\alpha + ( \mu + p ) u_\alpha,
   \nonumber \\
   & & S = 3 p, \quad
      \bar S^\alpha_\beta = \pi^{(3)\alpha}_{\;\;\;\;\;\beta},
   \label{ADM-variables}
\eea
where the intrinsic curvature $R^{(h)\alpha}_{\;\;\;\;\;\;\;\beta\gamma\delta}$
is a Riemann curvature based on $h_{\alpha\beta}$;
$\bar K$ is the sign of three-space curvature.
Thus, $\varphi$ is proportional to the perturbed three-space curvature
of the hypersurface normal to $n_a$.

{
\baselineskip 0pt
\addcontentsline{toc}{section}{References}

}


\begin{figure}[t]
   \centering
   \leavevmode
   \epsfysize=8cm
   \epsfbox{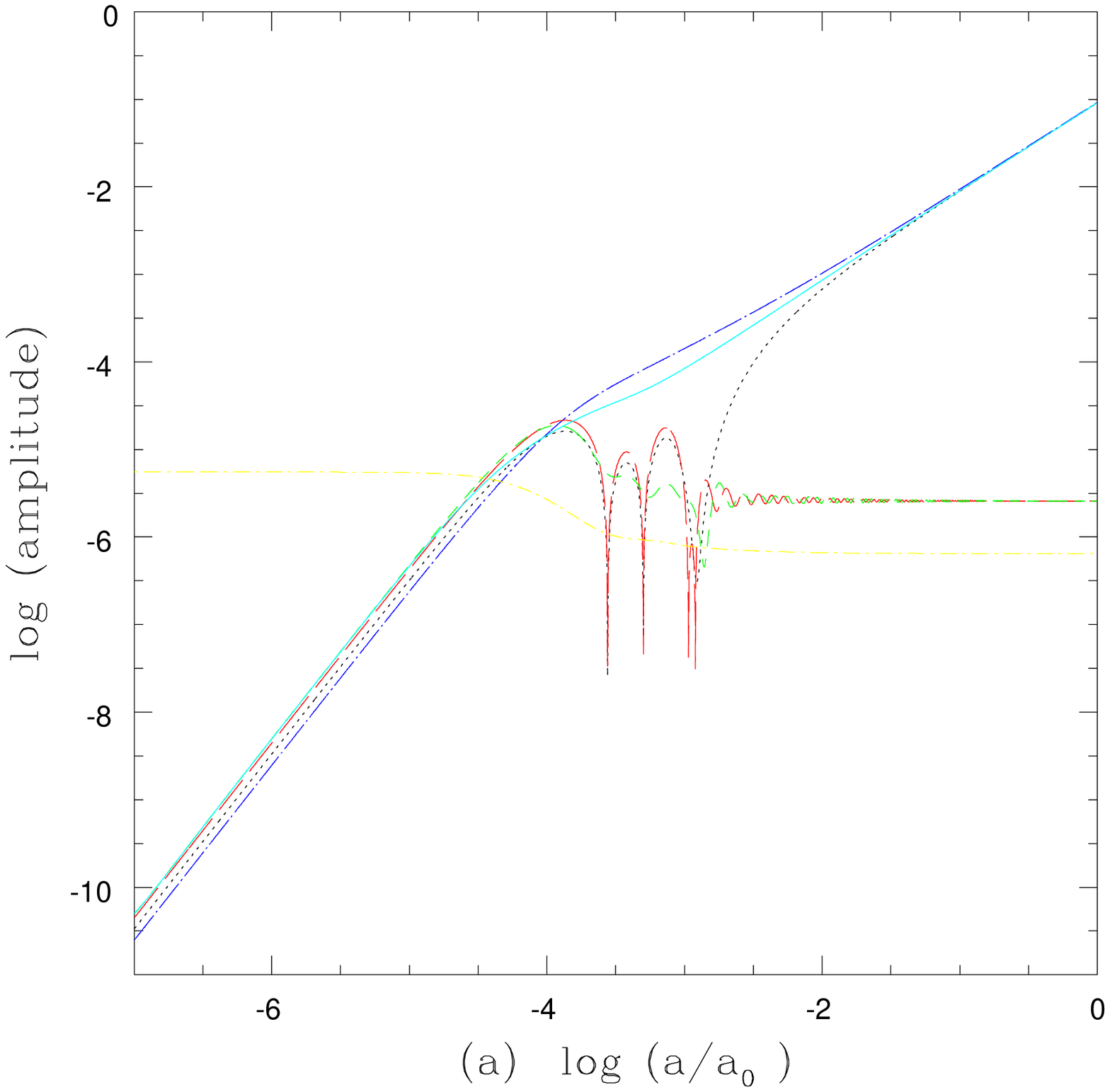}
   \nonumber
\end{figure}

\begin{figure}[t]
   \centering
   \leavevmode
   \epsfysize=8cm
   \epsfbox{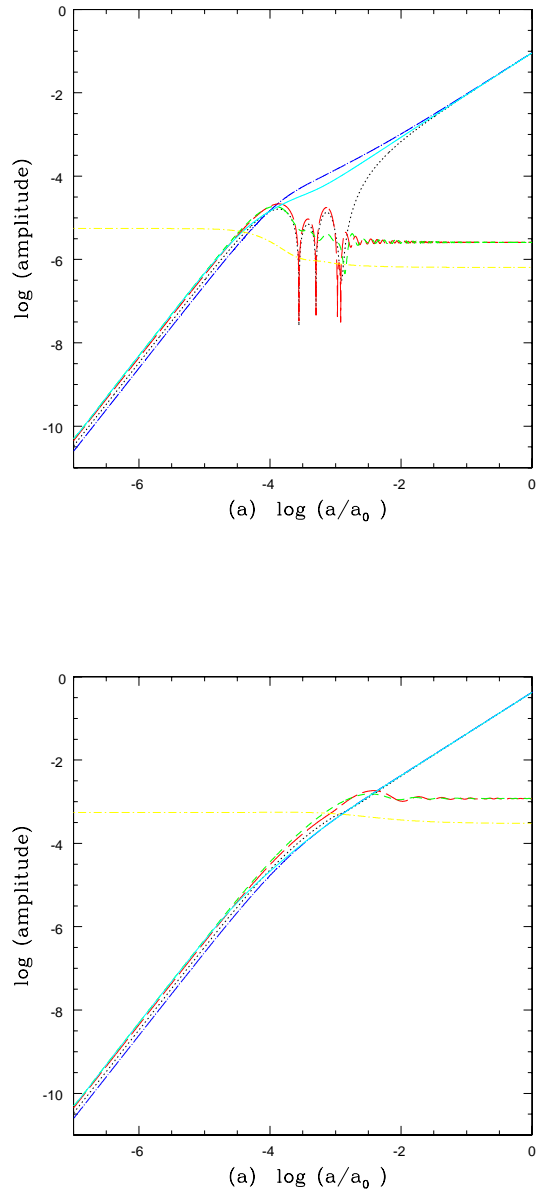}\\
   \caption[Fig_delta0.1]
   {\label{Fig_delta0.1}
We present the evolutions of the adiabatic density perturbations
in the corresponding comoving gauges $\delta_{(i)v_{(i)}}$
for several components.
$(i)$ includes the  baryon (dot, black), CDM (dot-long dash, blue),
photon (long dash, red), massless neutrino (short-dash, green), 
and massive neutrino (solid, cyan).
The two figures are (a) $k/a_0=0.1 Mpc^{-1}$ and  (b) $k/a_0=0.01 Mpc^{-1}$.
Also presented is the $\varphi_v$ (dot-short dash, yellow)
where $v$ is the collective fluid velocity.
The parameters are:
$h = 0.65$,
$\Omega_{b0} = 0.06$,
$\Omega_{C0} = 0.5$, 
$\Omega_{\gamma 0} = 5.85\times 10^{-5}$,
$\Omega_{\nu 0} = 3.99\times 10^{-5}$, and 
$\Omega_{\nu_m 0}= 0.44$.
We consider a flat background with vanishing $\Lambda$.
The absolute value of the vertical scale is arbitrary.
    }
   \label{fig:density1}
\end{figure}

\begin{figure}[t]
   \centering
   \leavevmode
   \epsfysize=8cm
   \epsfbox{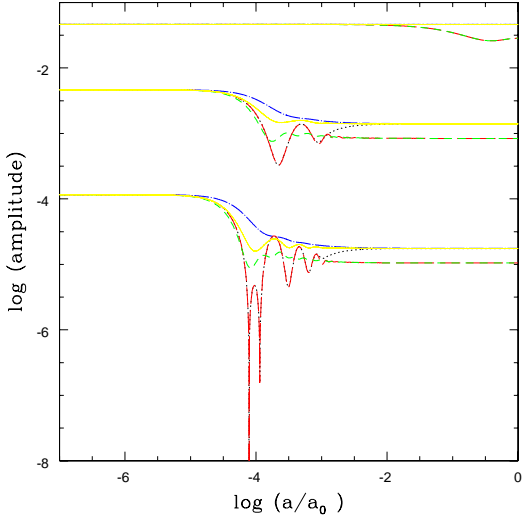}\\
   \caption[Fig_Phi1]
   {\label{Fig_Phi1}
We present the evolutions of $\varphi$ in various comoving gauge conditions
based on fixing the velocity variables of the components,
i.e., $\varphi_{v_{(i)}}$, and $\varphi_v$:
the baryon $\varphi_{v_{(b)}}$ (dot, black),
the CDM $\varphi_{v_{(C)}}$ (dot-long dash, blue),
the photon $\varphi_{v_{(\gamma)}}$ (long dash, red),
the massless neutrino  $\varphi_{v_{(\nu)}}$ (short dash, green),
and the one based on the collective velocity $\varphi_{v}$ 
(dot-short dash, yellow).
We consider three different scales: $k/a_0=0.001$ (top), $0.1$
and $0.2 Mpc^{-1}$ (bottom).
{}For $k/a_0=0.001 Mpc^{-1}$, the baryon, CDM, and the collective one are
overlapped (top),
and the photon and massless neutrino are overlapped (bottom).
In order to present the behaviors in three scales in one frame, we change the
absolute scale of the amplitude arbitrarily.
The parameters are:
$\Omega_{C0} = 0.94$,
$\Omega_{\nu_m 0} = 0$, 
and the other parameters are the same as in Fig. \ref{fig:density1}.
    }
   \label{fig:varphi1}
\end{figure}

\begin{figure}[t]
   \centering
   \leavevmode
   \epsfysize=8cm
   \epsfbox{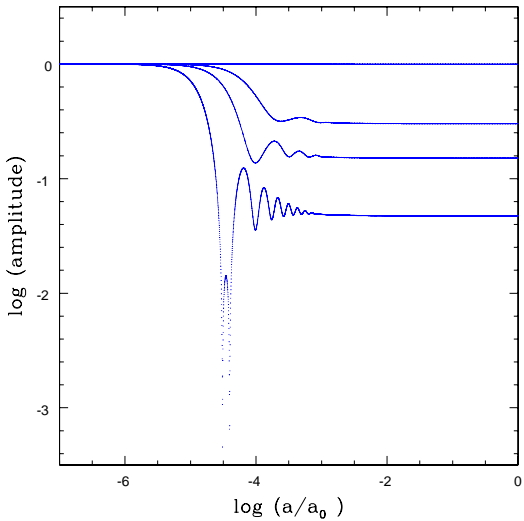}\\
   \caption[Fig_Phi2]
   {\label{Fig_Phi2}
The evolution of $\varphi_v$ for different scales:
$k/a_0= 0.0001$ (top), $0.001$, $0.01$, $0.2$, and $0.5 Mpc^{-1}$ (bottom).
The cases of $k/a_0=0.0001$ and $0.001 Mpc^{-1}$ are almost overlapped.
The parameters are the same as in Figure \ref{fig:varphi1}.
    }
   \label{fig:varphi2}
\end{figure}

\begin{figure}[t]
   \centering
   \leavevmode
   \epsfysize=8cm
   \epsfbox{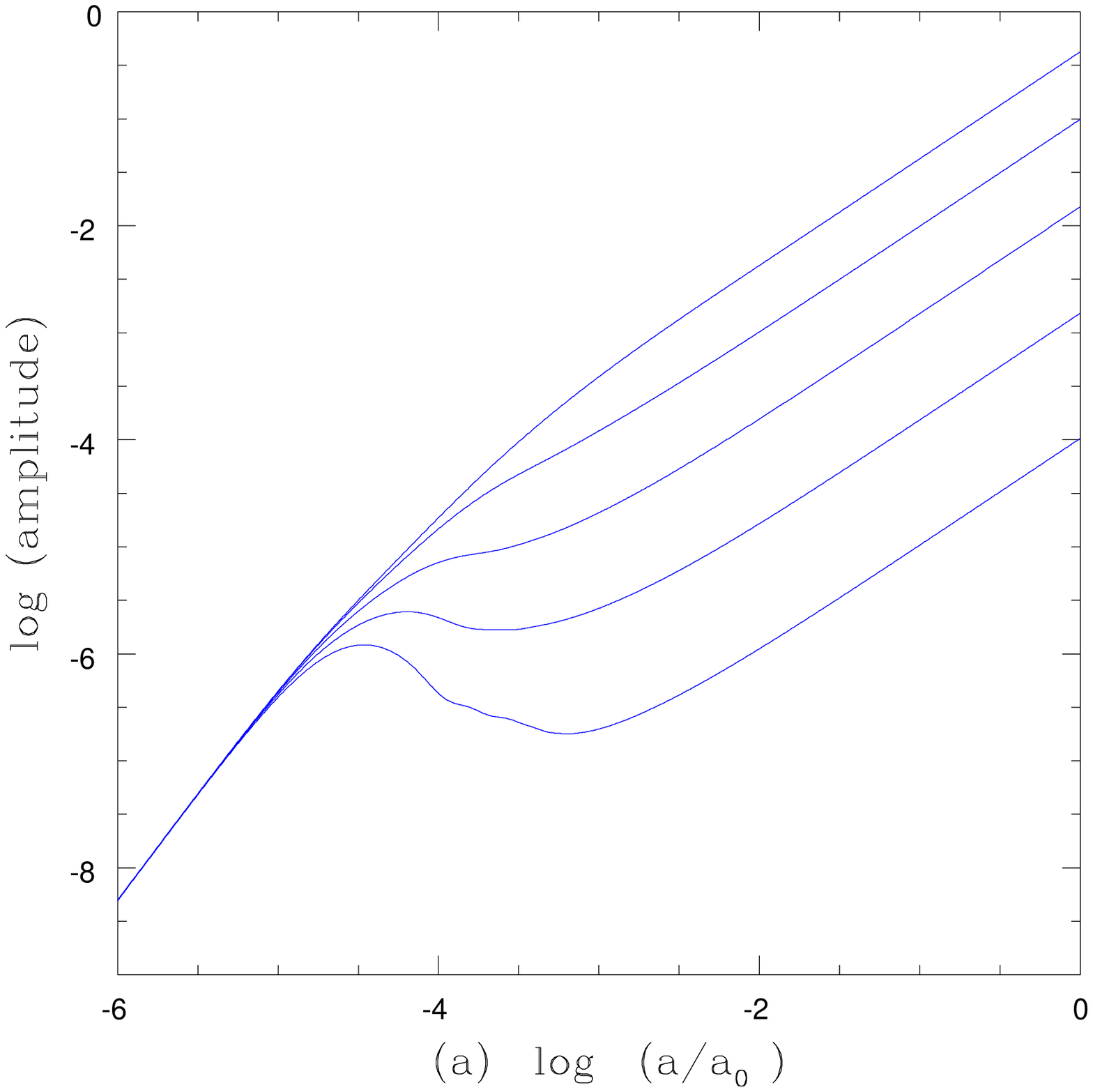}
   \nonumber
\end{figure}

\begin{figure}[t]
   \centering
   \leavevmode
   \epsfysize=8cm
   \epsfbox{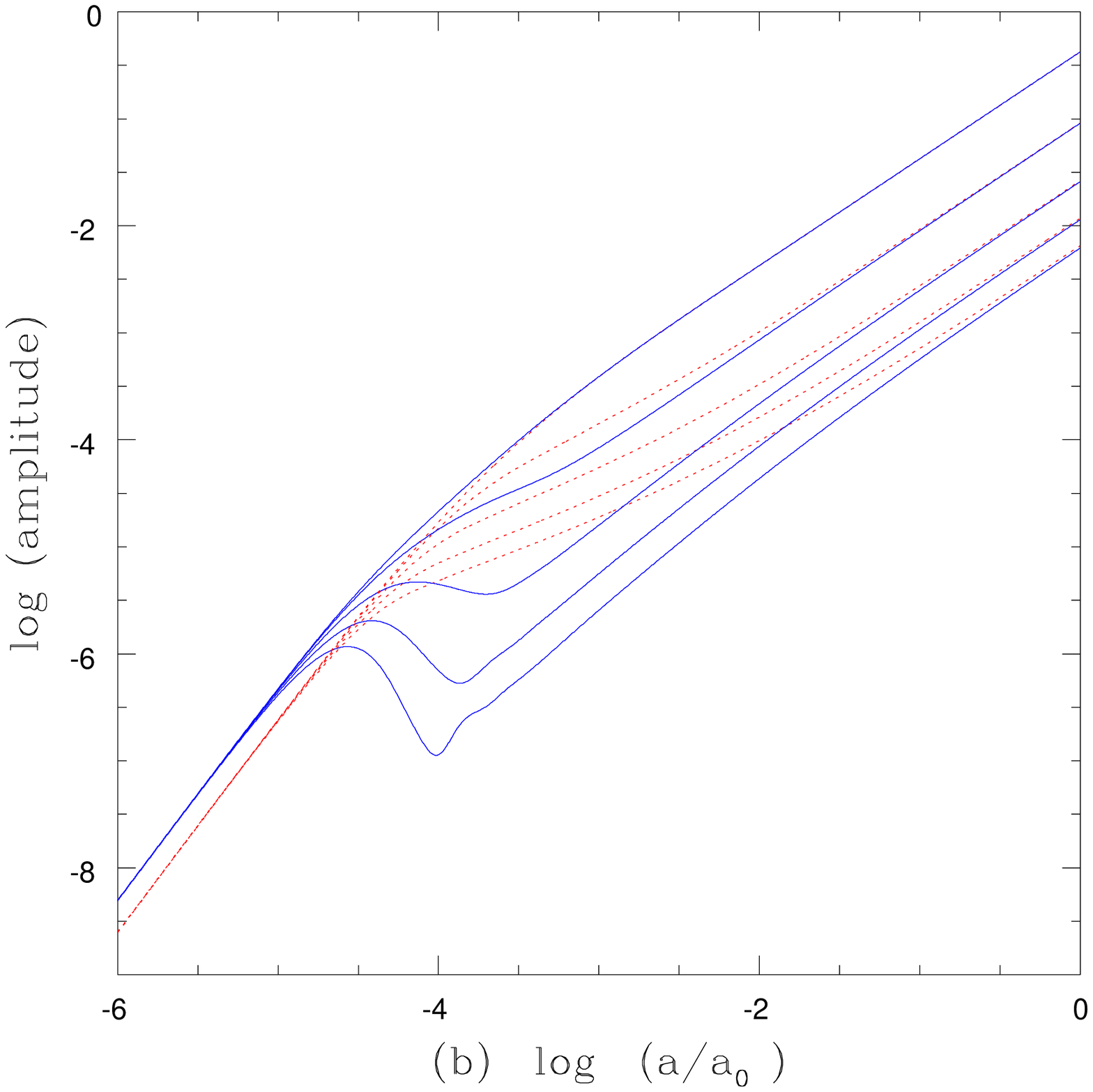}\\
   \caption[Figmassive1]
   {\label{Fig_massive1}
We present the evolution of density perturbation in the comoving gauge
of the massive neutrino, $\delta_{(\nu_m) v_{(\nu_m)}}$,
for different scales: $k/a_0 = 0.01$ (top), $0.1$, $0.2$, $0.3$ and
$0.4 Mpc^{-1}$ (bottom).
Figure (a) considers the massive neutrino dominated model with a parameter 
$\Omega_{\nu_m 0}=0.94$.
Figure (b) considers substantial amount of the CDM with parameters
$\Omega_{C0} = 0.5$ and 
$\Omega_{\nu_m 0}= 0.44$.
In Figure (b) we show the evolution of CDM
$\delta_{(C)v_{(C)}}$ (dotted, red) as well.
The other parameters are the same as in Fig. \ref{fig:varphi1}.
    }
   \label{fig:massive3}
\end{figure}

\begin{figure}[t]
   \centering
   \leavevmode
   \epsfysize=8cm
   \epsfbox{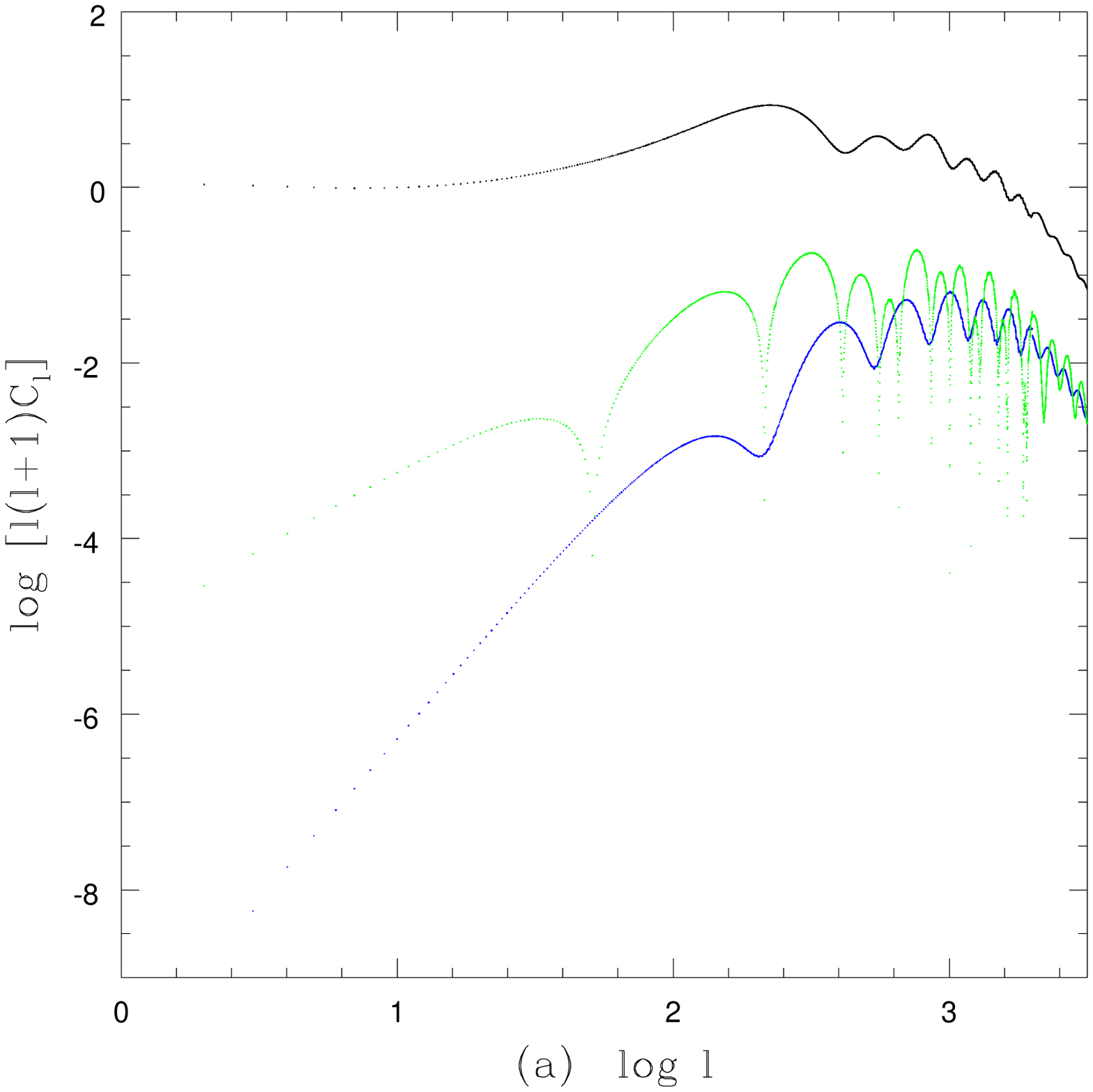}
   \nonumber
\end{figure}

\begin{figure}[t]
   \centering
   \leavevmode
   \epsfysize=8cm
   \epsfbox{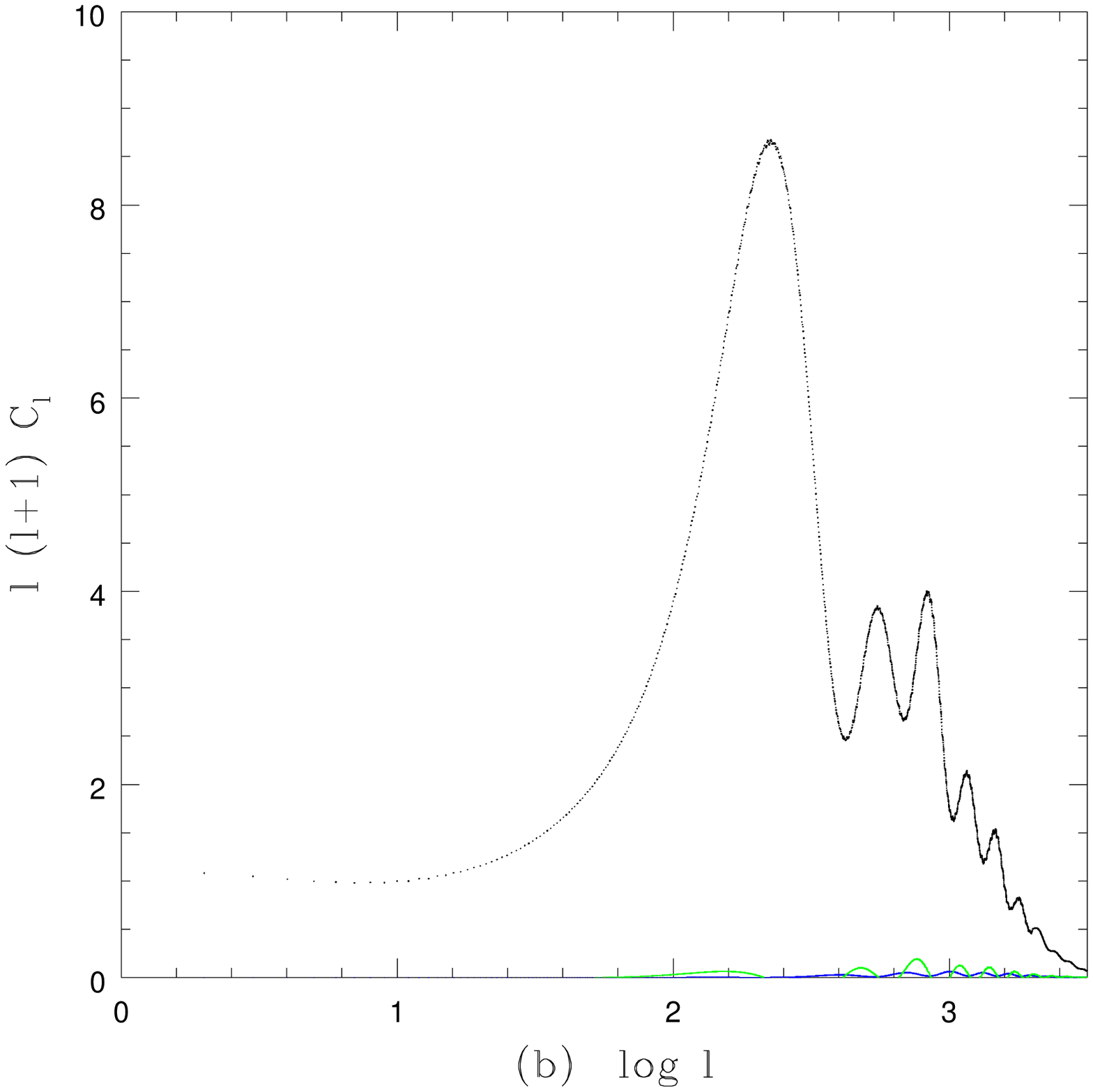}\\
   \caption[Fig_scalar1]
   {\label{Fig_scalar1}
We present the power spectra $\ell ( \ell + 1) C_\ell$
of the scalar-type perturbation:
the temperature $C_\ell^{\Theta \Theta}$ (top, black),
the polarization $C_\ell^{EE}$ (middle, green),
and the cross correlation $C_\ell^{\Theta E}$ (bottom, blue).
We take the adiabatic initial condition with the scale-invariant
($n_S = 1$) spectrum.
Figure (a) shows the spectra in logarithmic scale,
and (b) shows in real scale.
We normalize the spectra using
$\ell ( \ell + 1) C_\ell^{\Theta \Theta} = 1$ for $\ell =10$.
The parameters are:
$\Omega_{C0}=0.25$,
$\Omega_{\Lambda 0} = 0.69$,
$\Omega_{\nu_m 0}=0.$, and the other parameters are the same as 
in Fig. \ref{fig:density1}.
    }
   \label{fig:Cl}
\end{figure}

\begin{figure}[t]
   \centering
   \leavevmode
   \epsfysize=8cm
   \epsfbox{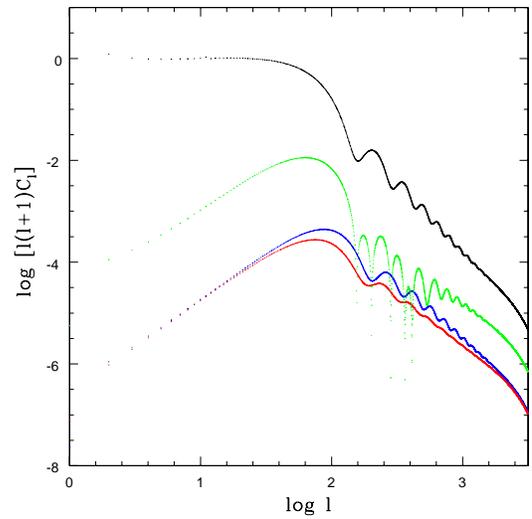}\\
   \caption[Fig_GW]
   {\label{Fig_GW}
We present the power spectra $\ell ( \ell + 1) C_\ell$
of the gravitational wave:
the temperature $C_\ell^{\Theta \Theta}$ (top, black),
the cross correlation $C_\ell^{\Theta E}$ (green),
and the polarizations $C_\ell^{EE}$ (blue)
and $C_\ell^{BB}$ (bottom, red).
As the initial condition we take the scale-invariant ($n_T = 0$)
spectrum and the solution with constant amplitude.
We normalize the spectra using
$\ell ( \ell + 1) C_\ell^{\Theta \Theta} = 1$ for $\ell =10$.
The parameters are the same as in Figure \ref{fig:Cl}.
    }
   \label{fig:ClGW}
\end{figure}

\end{document}